\pdfoutput=1
\documentclass[acmtog,final]{acmart}

\acmSubmissionID{2109}

\usepackage{booktabs}
\usepackage{threeparttable}
\usepackage{makecell}

\usepackage{tabularx}
\usepackage[ruled,linesnumbered]{algorithm2e}
\usepackage{multirow}
\usepackage{cleveref}

\usepackage{amsmath,amssymb}
\usepackage{tikz}

\usepackage[most]{tcolorbox}

\newtcolorbox{myshaded}{
  sharp corners,
  colback=gray!15,
  colframe=gray!15,
  boxrule=0pt,
  left=5pt,
  right=5pt,
  top=5pt,
  bottom=5pt,
  width=\linewidth,
  arc=0pt, outer arc=0pt
}

\usepackage{microtype}
\usepackage[ruled]{algorithm2e}

\SetAlFnt{\small}
\SetAlCapFnt{\small}
\SetAlCapNameFnt{\small}
\SetAlCapHSkip{0pt}

\setcopyright{cc}
\setcctype{by}
\acmJournal{TOG}
\acmYear{2026} \acmVolume{45} \acmNumber{6} \acmArticle{225}
\acmMonth{12} \acmDOI{10.1145/3842564}

\begin{document}
\title{Dynamic GPU Sparse Grids and Voxel-sequential Processing in NanoVDB\!}




\author{Eftychios Sifakis}
\orcid{0000-0001-5608-3085}

\affiliation{%
\institution{NVIDIA}
  \country{USA}
}
\email{esifakis@nvidia.com}
\affiliation{%
  \institution{University of Wisconsin-Madison}
  \country{USA}
}
\email{sifakis@cs.wisc.edu}

\author{Francis Williams}
\orcid{0000-0003-2189-881X}
\email{francis@fwilliams.info}
\affiliation{%
  \institution{NVIDIA}
  \country{USA}
}

\author{Christopher Horvath}
\orcid{0009-0002-7122-7117}
\email{chorvath@nvidia.com}
\affiliation{%
  \institution{NVIDIA}
  \country{USA}
}

\author{Rohan Sawhney}
\orcid{0000-0002-3661-1554}
\email{rsawhney@nvidia.com}
\affiliation{%
  \institution{NVIDIA}
  \country{USA}
}

\author{Mark Harris}
\orcid{0009-0008-4253-2917}
\email{mharris@nvidia.com}
\affiliation{%
  \institution{NVIDIA}
  \country{Australia}
}

\author{Jonathan Swartz}
\orcid{0000-0001-6320-2636}
\email{jswartz@nvidia.com}
\affiliation{%
  \institution{NVIDIA}
  \country{New Zealand}
}

\author{Ken Museth}
\orcid{0000-0002-9926-780X}
\email{kmuseth@nvidia.com}
\affiliation{%
  \institution{NVIDIA}
  \country{USA}
}

\renewcommand{\shortauthors}{Sifakis et al.}
\begin{teaserfigure}
  \includegraphics[width=\textwidth]{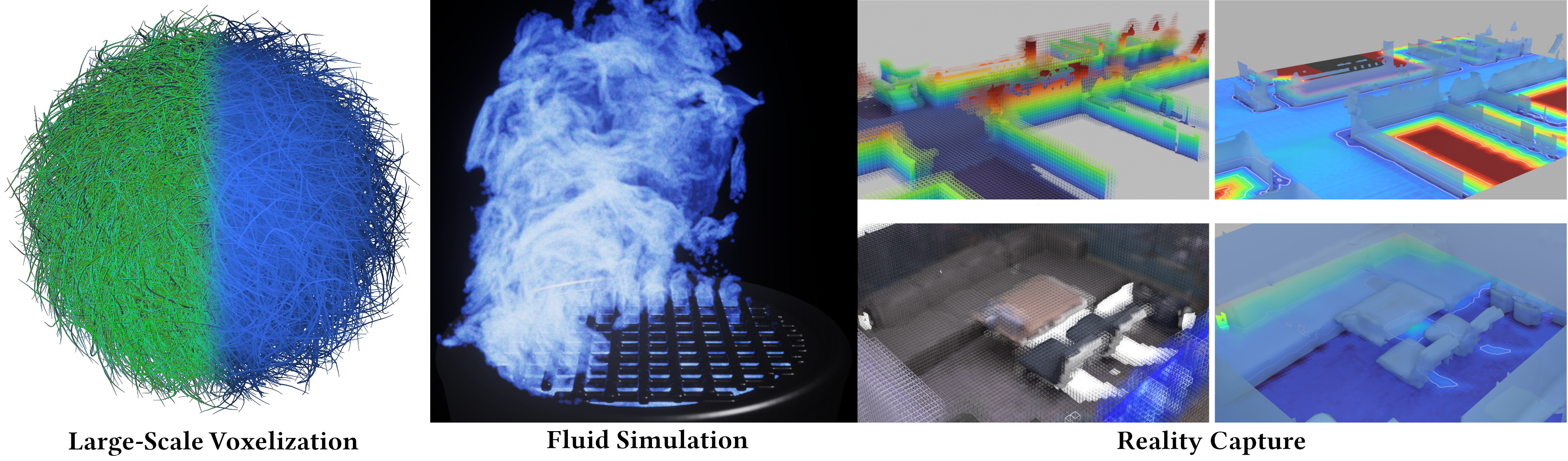}
  \caption{
  We demonstrate our enhancements to NanoVDB in three diverse applications that rely on dynamic sparse grids. From left to right: \textbf{Large-Scale Voxelization} of a 2.8M-triangle hairball mesh \citep{McGuire2017Data} (left hemisphere in green), our \texttt{MeshToGrid} implementation generates a sparse index grid with $9.88$B active voxels and UDF values stored in a sidecar (right hemisphere in blue) in $3.24$s;
  \textbf{Fluid Simulation} of fire using the ghost fluid method \cite{10.1145/566654.566643, NGUYEN200171, 10.1145/1275808.1276436} to track the combustion flame front using a truncated signed distance field;
  a \textbf{Reality Capture} processing pipeline which fuses a TSDF surface mesh, on the left, and derives a Euclidean signed distance
field (ESDF), on the right, to facilitate collision checking and path planning from an input LiDAR sweep or a sequence of depth images. Hardware details in Appendix \ref{appendix:hardware-config}.
    }
  \label{fig:teaser}
\end{teaserfigure}

\begin{abstract}

We introduce two fundamental advances to the NanoVDB
data structure: dynamic GPU-native topological updates, and a voxel-sequential access paradigm that dramatically accelerates point-wise and stencil-based operators on sparse grids. NanoVDB's linearized memory layout has delivered strong GPU performance, and the Index-Grid concept introduced in the fVDB framework \cite{Williams:2024:fVDB} provided topology/data separation and optimal compactness of storage in sparse machine learning workloads. However, all these prior iterations of NanoVDB assume a static topology that has precluded use of this data structure in simulation and dynamic geometric processes. We eliminate this limitation by introducing a suite of dynamic topology operators -- morphological dilation, grid merging, pruning, and cross-resolution coarsening and refinement. These operators use bit-parallel algorithms to reconstruct grids containing hundreds of millions of voxels in just a few milliseconds.

We also introduce a voxel-sequential access paradigm and data structure for GPU kernels operating on sparse data. We stream active voxels as a flat linear sequence, minimizing or eliminating the need for tree traversal in the process. This achieves compute density and regularity that allow even complex stencil-based operators to approach peak memory bandwidth, with minimal storage overhead. Together, these contributions remove the read-only constraint from NanoVDB, establishing it as a high-performance primary representation for dynamic GPU simulation and geometry manipulation. We evaluate the efficacy of our newly introduced operators in benchmarks sourced from applications such as 3D reconstruction, fluid simulation, and large-scale mesh voxelization.
\end{abstract}

\maketitle
\section{Introduction}

The need to store and manipulate high-resolution, sparse volumetric data is ubiquitous across the spectrum of modern spatial computing. From high-fidelity fluid simulations in visual effects to massive datasets in medical imaging and neural radiance fields, a few recurring technical challenges emerge prominently: \emph{efficiency} in memory utilization, \emph{scalability} in computational throughput and accommodation of larger domain sizes, and the necessity of \emph{sparsity} in representing irregular 3D domains. The OpenVDB data structure \cite{Museth:2013} and its various derivatives were born out of these challenges, emerging as the industry standard for encoding sparse grid-embedded data. By providing a framework to represent virtually unbounded Cartesian domains with sparse occupancy, OpenVDB enjoys a rich ecosystem of operators and broad adoption in digital content creation tools. More recently, NanoVDB \cite{Museth:2021} has extended this functionality to the GPU with emphasis on implicit geometry modeling and real-time rendering tasks, while the fVDB framework \cite{Williams:2024:fVDB} further extended the scope of NanoVDB and demonstrated its efficacy in sparse machine learning workloads.  NanoVDB has seen industry-wide adoption, powering tools such as Blender’s Cycles rendering engine, Houdini’s Pyro, Vellum and MPM solvers, Autodesk's Arnold, Chaos' V-Ray, and Pixar's RenderMan.

Despite its success, the VDB ecosystem -- as originally materialized in OpenVDB -- was fundamentally conceived for the CPU, relying on fine-grained heap allocation and pointer-heavy nodes, accelerated by means of a software cache implemented in \emph{ValueAccessor} objects. Although this design allows for the effortless topological change required by dynamic simulation, it is not the most natural fit for the massively parallel, throughput-oriented architecture of the GPU. NanoVDB \cite{Museth:2021} was introduced to facilitate the transition of VDB grids to GPU. Originally intended for read-only applications, e.g. random access tasks like ray-tracing and collision detection, NanoVDB forgoes heap allocation and pointer-based bookkeeping by adopting a linearized memory layout with relative addressing, specifically tailoring the structure for SIMT execution.

In a phenomenon that  should not surprise us given the precedent of other high-throughput systems, NanoVDB's narrower scope -- sacrificing OpenVDB's topological malleability for a compact, pre-baked buffer -- results in a structure that is not only more portable but significantly faster for streaming, sampling, and rendering tasks. Consequently, OpenVDB remains the tool of choice for dynamic, evolving topologies at the cost of peak performance, while NanoVDB serves as the expert subset for GPU-accelerated workloads, limited by its static, read-only topology. So far, no published research has seriously challenged this limitation, leaving developers with an untenable choice: the dynamic topological versatility of the CPU (OpenVDB) or the read-only speed of its GPU sibling (NanoVDB).

In this paper, we resolve this tension by introducing a framework for the dynamic evolution of NanoVDB sparse volumes directly on the GPU. We demonstrate that static topology is not an inherent trait of the NanoVDB data structure, but rather a limitation of the algorithms previously used to maintain it. Our work specifically aims to be synergistic with NanoVDB index-grids \cite{Williams:2024:fVDB}, the latest feature iteration of the data structure that explicitly separates the active voxel set topology from the data payload stored in it, and also offers the highest compactness of storage of all VDB variants. Our tools target the
rapid reconstruction of the tree's hierarchical backbone independently of the underlying data payload. We adopt a bulk-synchronous manipulation paradigm where topological changes are enacted collectively in discrete epochs, and enable a suite of operators -- including morphological dilation, grid merging, pruning, and multi-resolution refinement -- that can reconstruct trees with hundreds of millions of voxels in a few milliseconds.

With topological manipulation now handled efficiently on the GPU, the remaining fundamental concern for sparse simulation codes is the performance of streaming and stencil operations. \emph{Accessor} objects -- originating in OpenVDB and carried into NanoVDB -- are the standard paradigm for data access in the VDB ecosystem. They act as software caches that hide the cost of tree traversal, if the locations being accessed are spatially coherent. They are also available on the GPU (in NanoVDB) although their performance benefit may be hampered at times due to thread divergence when an accessor needs to update its local cache after a miss. Another challenge stems from the fact that on the GPU, the natural way to dispatch parallel threads is to assign a threadblock (e.g. 512 CUDA threads) to process all voxels of a leaf node in parallel, leading to load imbalance if threads are wasted being assigned to inactive voxels. Although accessor-based random access still has its place (especially in ray tracing and collision processing workloads), we propose an alternative data traversal paradigm, which we label \emph{voxel-sequential access} that is optimized for stencil and streaming operations on the GPU. Within this paradigm, we stream active voxels as a flat linear sequence, regardless of the leaf nodes they straddle along this ordering. We facilitate this via a new \emph{VoxelBlockManager} data structure that opens the path to streaming and stencil operations at optimal compute density, and can absorb the cost of tree traversal via enhanced regularity and warp-synchronous operation.

Our specific contributions are:
\begin{itemize}
    \item \textbf{Bulk-Synchronous Topological Evolution:} A paradigm for GPU- and VDB-native topological changes---including merging, pruning, and refinement---on linearized index-grids. We argue this is the most fitting abstraction of \emph{dynamic} topology evolution in GPU-focused workloads. We demonstrate bit-parallel algorithms that reconstruct complex topologies with hundred million voxels in milliseconds.

    \item \textbf{Injection and Sidecar Operators:} A suite of operators for the efficient integration and evolution of auxiliary \emph{sidecar} payloads alongside the primary topology.

    \item \textbf{The Voxel-Sequential Access Paradigm:} The \emph{VoxelBlockManager} (VBM) data structure, which replaces legacy hierarchical traversal with a linear streaming pattern. This eliminates thread divergence and load imbalance, significantly boosting efficiency on stencil-oriented workloads.

    \item \textbf{High-Performance Validation:} We demonstrate the efficacy of our framework additions in microbenchmarks sourced from real-world workloads, including 3D geometry reconstruction, interface tracking in fluids simulation, and high-resolution 3D mesh voxelization.
\end{itemize}

\begin{figure}[t]
\centering
\includegraphics[width=\columnwidth]{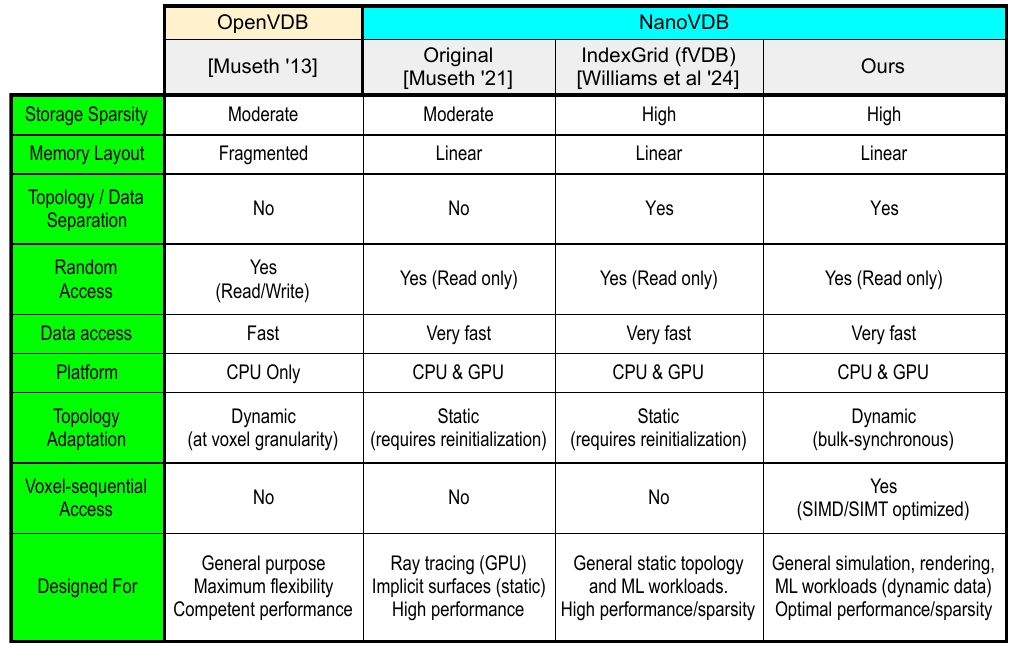}
\caption{\textbf{Feature comparison of VDB iterations} Our proposed enhancements combine the highly compact storage of NanoVDB Index-Grids, the GPU-compatibility of all NanoVDB iterations, support for GPU-native topology modification operations, and a new voxel-sequential access paradigm}
\label{fig:vdb-comparison}
\end{figure}
\section{Background and Related Work}\label{sec:background-related}

We will start with an overview of the evolution of the data structures in the VDB ecosystem, and the implications that the design decisions in every iteration brought about. Figure \ref{fig:vdb-comparison} highlights some of the key architectural differences, which are further detailed below.

\textbf{OpenVDB} \cite{Museth:2013} introduced the core concept of the VDB tree structure: a hierarchical tree-based representation with fixed depth but aggressive and varying branching factors. At the coarsest scale, the sparse 3D voxel domain is covered by aligned macroscopic \emph{root tiles} (nominally spanning $4096^3$ voxels), which are maintained by the root node in a linear list or a hashtable.

If a root tile requires refinement rather than carrying a constant value, an \emph{upper node} is associated with it, spanning the same area and holding the branch of the tree that refines it. Upper nodes are subdivided into voxel partitions of size $128^3$ voxels (so called \emph{lower nodes}) and those into smaller aligned partitions of $8^3$ (\emph{leaf nodes}).

A leaf node densely allocates a buffer (on the heap) sufficient to store all $8\times 8\times 8$ values within its span, and a bit-encoded mask indicating an active/inactive state for each voxel in its domain. OpenVDB strives to present a usage model that is as permissive as a hypothetical, infinite-extent dense grid would be: random read/write access is supported to arbitrary locations in the Cartesian domain, with the data structure performing on-demand allocation to create storage for any location that was not previously provisioned for.

The fundamental abstraction that OpenVDB uses to mitigate the (substantial) worst-case performance cost of random access is the concept of \emph{accessors}, cf.~\texttt{ValueAccessor} in \cite{Museth:2013}. An accessor is a tool for read/write access at arbitrary coordinates, which seeks to accelerate access cost if the read/write requests are made for spatially coherent locations. Specifically, if repeated access requests are made for voxels that fall within the same leaf node, a software cache that the accessor incorporates will completely bypass tree traversal and provide fast access to the requested value. That is, an accessor allows for bottom-up vs top-down tree traversal, optimally avoiding traversal altogether. However, the highly permissive use model of OpenVDB limits its potential for aggressive optimization: the data structure is clearly targeting CPU-only usage, as accessors are prone to warp divergence (they need to modify their internal cache even if they access data for read purposes only), and the allocate-on-insertion model for topology updates is very challenging to reconcile with the typical GPU/CUDA programming paradigm.

\begin{figure}[t]
\centering
\includegraphics[width=\columnwidth]{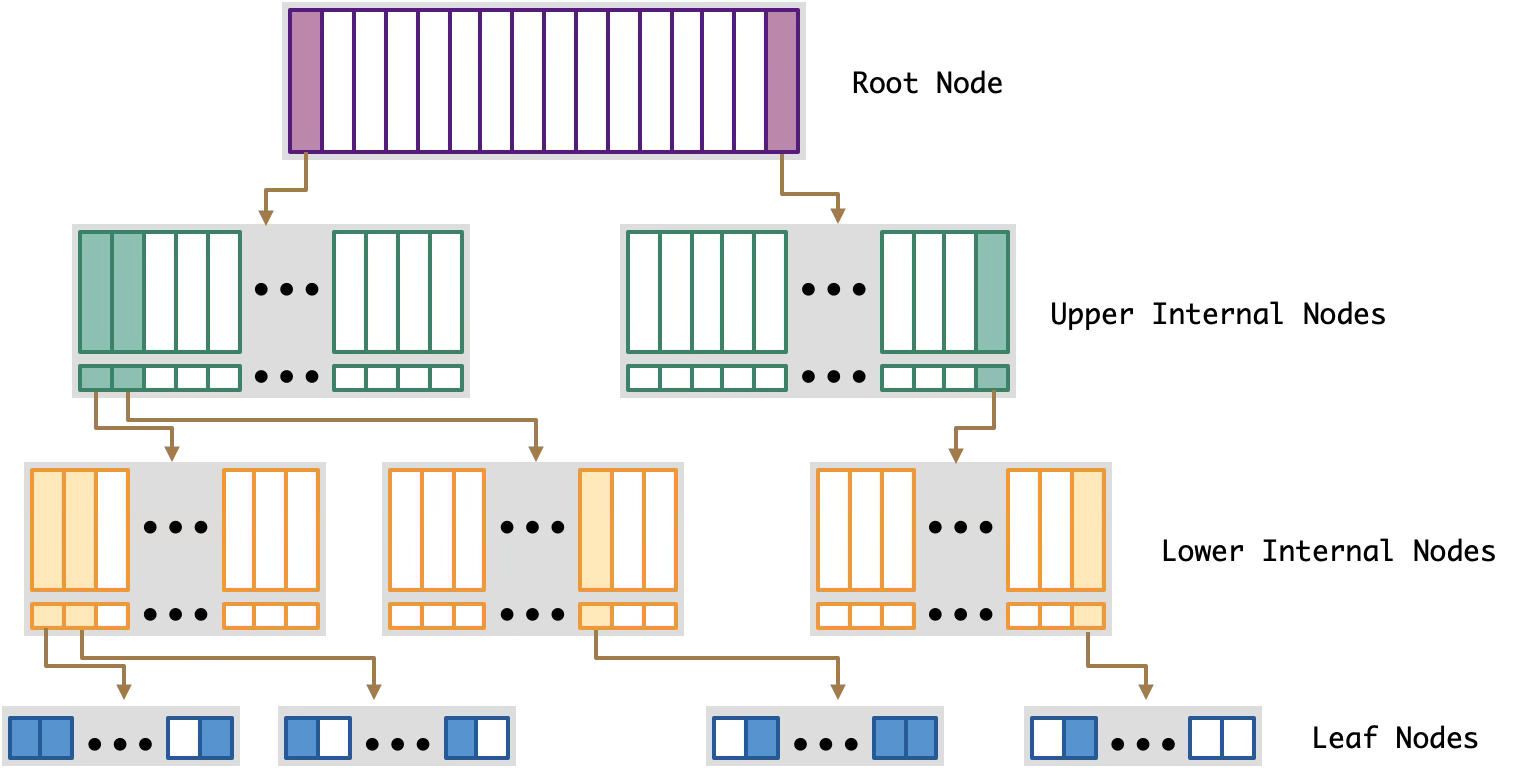}

\medskip
 \setlength{\tabcolsep}{4pt}
 \begin{tabular}{lcccc}
   \toprule
   Name & $\mathrm{LOG2DIM}$ & Branching & Voxel span\\
   \midrule
   Root  & --- & $n_\mathrm{Tiles}$ & unbounded  \\
   Upper Internal & 5 & $32^3 = 32{,}768$ & $4096^3$  \\
   Lower Internal & 4 & $16^3 = 4{,}096$  & $128^3$   \\
   Leaf  & 3 & $8^3 = 512$        & $8^3$       \\
   \bottomrule
 \end{tabular}
\caption{\textbf{Branching factor and voxel span} for each level of the NanoVDB tree ($\mathrm{LOG2DIM} = 5,4,3$). Each node represents a span of $2^{3\cdot\mathrm{LOG2DIM}}$ voxels, at the leaf level. The root holds an unbounded, hash-keyed list of $n_\mathrm{Tiles}$ tiles.}
\label{fig:nvdb-illustration}

\end{figure}

\textbf{NanoVDB} \cite{Museth:2021} was introduced as a narrowly-scoped subset of OpenVDB that was expressly aimed at maximizing portability, supporting use in both CPU and GPU, x86 and ARM platforms, and even within shading languages. NanoVDB froze the geometric sizes of the tree nodes from being tunable at compile-time in OpenVDB, to a standardized 4096/128/8 node size progression as illustrated in Figure \ref{fig:nvdb-illustration}, and replaced the \emph{allocation-on-access} model of OpenVDB with a strict \emph{allocation-on-initialization} without support for changing the grid topology after it was created. Leaf nodes and their data would be allocated in a contiguous, linear storage span, rather than the fragmented heap-based allocation that OpenVDB required. NanoVDB is more efficient in (spatially coherent) random access than OpenVDB on the CPU, due to the fact that NanoVDB uses linearized, cache- and prefetch-friendly storage. Of course, in this initial iteration of NanoVDB, its linear memory model came at the expense of no support for performing edits to the grid topology (modifying the active voxel set) without requiring rebuilding the grid from scratch.

\textbf{Index-Grids} \cite{Williams:2024:fVDB} were a crucial addition to NanoVDB, introduced to facilitate support for machine learning workloads in the fVDB framework. In an Index-Grid, the value associated with each active voxel is not the actual data ``payload'', but an integer index into an auxiliary linear storage, referred to as a \emph{sidecar}. This topology/data separation provided a number of substantial benefits. First, it is trivial to store scalar fields corresponding to distinct quantities associated with each voxel (e.g. density, pressure, level set value, etc in a fluid simulation), while reusing the same topology, but distinct ``sidecar''. Second, it provides the maximum potential for sparsity, as the sidecar only needs to allocate space for \emph{active} coordinates, at voxel granularity; in contrast, storage for the prior NanoVDB iterations had to be allocated for all $8^3$ voxels of every leaf node, regardless of how many of such voxels are active. Finally, the actual linear index associated with every active voxel coordinate does not need to be explicitly stored, but computed on-the-fly, with minimal storage footprint (typically less than 2 bits/voxel for grids with approximately $50\%$ leaf-level occupancy). Still, in this latest iteration as presented in fVDB \cite{Williams:2024:fVDB} NanoVDB (including IndexGrids) remained a read-only structure after initialization. Our work closes exactly this feature gap, describing a set of GPU-native dynamic topology manipulation operators.

At this point, we should clarify (and defend) the characterization of such operators as ``dynamic'', in contrast with the context that this term is used elsewhere in the VDB ecosystem. OpenVDB supports a truly laissez-faire paradigm of topology updates, where an individual voxel can be introduced or removed from the active set at any time. We argue that such an extremely permissive model of ad-hoc topology manipulation is not how typical high-throughput (and at the same time thread-safe and deterministic) simulation and geometry processing workloads ought to operate. OpenVDB topology updates are not thread-safe, and must be guarded in use-specific ways to ensure that each execution thread doesn't attempt to consume voxels that were asynchronously created by another. We advocate, especially in light of GPU-oriented workloads, for a bulk-synchronous model of topological adaptation, where the topological change is prescribed holistically for the entire domain (e.g. a morphological dilation, or a topological union of two grids), and enacted as one synchronous operation. Within this framework, we defend the ``dynamic'' label for our update operators in the sense that (a) they are GPU-native and highly optimized, and (b) they operate natively within the VDB representation. The one concession we commit to is the need to retain in memory both the ``before'' and ``after'' version of the VDB topology, while the operation takes place, although the absolute storage cost of this is small, due to the topology/data separation of NanoVDB Index-Grids.

\paragraph{Sparse grids for level set methods and volumetric modeling.}
Sparse volumetric grids have a long history in computer graphics, particularly in applications such as implicit modeling, fluid simulation, and level set methods. Because this body of literature is too extensive to cover in its entirety, we restrict our focus to data structures designed for narrow-band level set simulations -- a central application of the current work -- and exclude VDB-based approaches discussed above. Existing methods generally fall into three categories: hierarchical, blocked, or compressed representations. Hierarchical methods typically utilize height-balanced octrees \cite{STRAIN1999616, MIN2004368, 10.1145/1015706.1015745} to adaptively refine the grid near the interface; however, these structures often suffer from the computational overhead associated with deep tree traversals and pointer chasing. To address cache locality, blocked or tiled schemes, eg.~\cite{1250357}, group voxels into contiguous memory blocks, leveraging hardware-friendly layouts such as those found in SPGrids ~\cite{10.1145/2661229.2661269} and the more recent Multiresolution Sparse Block Grids (MSBG) ~\cite{10.1145/3730854}.
Beyond its memory layout, SPGrid also demonstrates that a pyramid of sparsely populated uniform grids can support adaptive, multi-resolution stencil work without explicit tree traversal -- the same coupling of resolution change and stencil computation that our coarsening and refinement operators are intended to support.
Conversely, compressed representations such as the Dynamic Tubular Grid (DT-Grid) \cite{DTGrid} and Hierarchical Run-Length Encoding (H-RLE) \cite{10.1145/1122501.1122508} generalize compressed-row-storage and RLE to 3D. While these formats provide exceptional memory efficiency for static volumes, they incur significant re-structuring costs when the topology is dynamic and requires frequent updates. Finally, out-of-core approaches ~\cite{10.1145/1289603.1289607} enable streaming computing of extremely large level sets, however their performance is fundamentally bottlenecked by I/O bandwidth and the latency of disk-to-CPU data transfers and as such cannot match the computational speeds of GPU approaches like ours.

\paragraph{Block-sparse hash grids for volumetric reconstruction.}
Outside the VDB ecosystem, the dominant sparse-volumetric representation in real-time 3D reconstruction is the \emph{block-sparse hash grid}: integer block coordinates (typically $8^3$ voxels per block) are inserted into a parallel hash map, and each occupied block heap-allocates a dense voxel payload. \citet{newcombe2011kinectfusion}'s KinectFusion popularised dense volumetric TSDF fusion from depth streams, and \citet{niessner2013hashing} replaced the dense grid with a GPU hash map to remove the bounded-domain restriction. This block-hash recipe has since been adopted across the field: Voxblox~\cite{oleynikova2017voxblox} extends it to incremental Euclidean signed distance fields for MAV planning, nvblox~\cite{millane2024nvblox} ports the same recipe onto the GPU, and Open3D~\cite{zhou2018open3d} ships a block-hash TSDF backend built on ASH~\cite{dong2023ash}, a modern GPU spatial-hashing framework. Within the VDB family itself, VDBFusion~\cite{vizzo2022vdbfusion} demonstrates that an OpenVDB-backed TSDF pipeline reaches comparable quality without committing to a specific block resolution at compile time. The octree alternative pioneered by OctoMap~\cite{hornung2013octomap} remains widely deployed in robotics for occupancy mapping. All of these systems are built around a static fusion-and-query loop --- they grow topology as new observations arrive but do not support the kind of bulk topological rewriting (dilation, union, pruning, refinement) that fluid simulation, dynamic-scene mapping, and geometry processing demand. Section~\ref{sec:reality_capture} compares directly against nvblox, VDBFusion, and Open3D on the same reconstruction benchmarks.

\paragraph{Sparse-voxel deep learning frameworks.}
A parallel line of work targets sparse 3D learning rather than reconstruction. Minkowski Engine~\cite{choy2019minkowski} and spconv~\cite{spconv2022} encode active coordinates in a hash table and run sparse convolutions over the resulting key-value structure; TorchSparse~\cite{tang2022torchsparse} and its successor TorchSparse++~\cite{tang2023torchsparsepp} extend this design with kernel-fusion and tile-sparsity optimisations. O-CNN~\cite{wang2017ocnn} and Kaolin~\cite{jatavallabhula2019kaolin} instead embed sparse data in an octree, while Taichi~\cite{hu2019taichi} provides a more general DSL for spatially-sparse compute that has been used both for simulation and for learning. NeuralVDB~\cite{kim2022neuralvdb} explores neural compression of VDB topology, complementary to the explicit dynamic operators we introduce here. Compared with these flat hash-keyed representations, the VDB hierarchy used by fVDB~\cite{Williams:2024:fVDB} retains spatial locality at every level of the tree, which our work exploits to make topological updates --- bit-parallel at the leaf, hash-based at the root --- both deterministic and fast.

\paragraph{Implicit and generative alternatives.}
Recent neural representations sidestep explicit sparse grids by encoding geometry in network weights and learned feature volumes. Instant-NGP~\cite{mueller2022instantngp} stores trainable features in a multiresolution hash table whose collisions are disambiguated by the downstream MLP --- an approach that is highly efficient for fitting but does not expose an explicit topology that downstream geometric operators can act on. Closer to our setting, NKSR~\cite{huang2023nksr} demonstrates large-scale neural surface reconstruction from sparse oriented points using a VDB-backed kernel basis, and XCube~\cite{ren2024xcube} performs voxel-hierarchy diffusion directly on VDB grids to generate $1024^3$ scenes in a single feed-forward pass. Both rely on fVDB for their voxel topology, and both would benefit directly from the dynamic topology operators we develop here: NKSR's basis support could be incrementally refined as new observations arrive, and XCube's coarse-to-fine generation could write its progressively finer levels through our refinement operator rather than rebuilding the grid from scratch.

\paragraph{Resolution-selection criteria}
A complementary line of work focuses on how resolution should adapt. In volumetric reconstruction the criterion is typically derived from an observation model rather than from geometry alone: \citet{vespa2019} select the integration scale within an octree from the effective sensor resolution at the observed depth, while \citet{derebotti2025} adapt voxel size to the local variance of SDF observations in a flat multi-resolution hash. Both evaluate the criterion per block and apply it incrementally, as observations arrive during fusion. Criteria of this kind are the natural drivers for the coarsening and refinement operators we introduce in \cref{sec:topology}, which supply the corresponding mechanism on the GPU: such a predicate, evaluated over the active set, becomes the input mask of a coarsening or refinement epoch, which the operator then enacts as a single bulk rewrite of the topology. We deliberately keep decision and mechanism separable -- the same primitives serve a sensor-driven, a variance-driven, or a purely geometric criterion without modification.
{}
\section{Dynamic Topology Evolution} \label{sec:topology}

 We propose a bulk-synchronous model for evolving the topological domain of active voxels in a NanoVDB sparse grid. Topological adaptation follows three steps: (a) topology changes (where and by what rule) are \emph{decided} collectively; (b) prescribed changes are \emph{materialized} in bulk, resulting in a new, valid topology and data state; and (c) computation resumes on this \emph{new epoch} of the grid.

We adopt this epoch-based paradigm as the foundation of our dynamic topology updates, providing five operators for evolving the hierarchical tree structure: (1) Morphological \emph{dilation} expands the active voxel set into its immediate neighborhood; (2) \emph{Pruning} removes voxels based on an input mask; (3) Merging computes a Boolean union of two Index-Grids; (4) Coarsening downsamples by a structured 2:1 ratio on each axis (8:1 volumetric reduction); and (5) Refinement, the adjoint of coarsening, upsamples by the same 2:1 ratio.

These operators compose into macroscopic, domain-specific workflows. For example, a typical level set narrow-band update is recast as three discrete steps: a speculative dilation to accommodate the anticipated motion of the interface; a fixed-topology numerical update of the level set values; and a pruning pass to re-establish the narrow-band limits and discard voxels that have moved beyond the region of interest. These operators are designed specifically for \emph{Index-Grids} \cite{Williams:2024:fVDB}, which only store the topology structure of the grid, leaving the data payload to a separate linearized sidecar. We first describe the common skeleton (and individual details) for these operators as they pertain to the rebuilding of the topology-only grid itself; operators that update the sidecar storage in response to topological changes are discussed in section \ref{sec:sidecar-ops}. The applications of \cref{sec:reality_capture} additionally use the prior \texttt{voxelsToGrid} primitive of \citet{Williams:2024:fVDB}, which rasterizes a set of integer coordinates into a fresh Index-Grid; we treat it as a black-box building block.

\begin{figure*}[t]
\centering
\includegraphics[width=\textwidth]{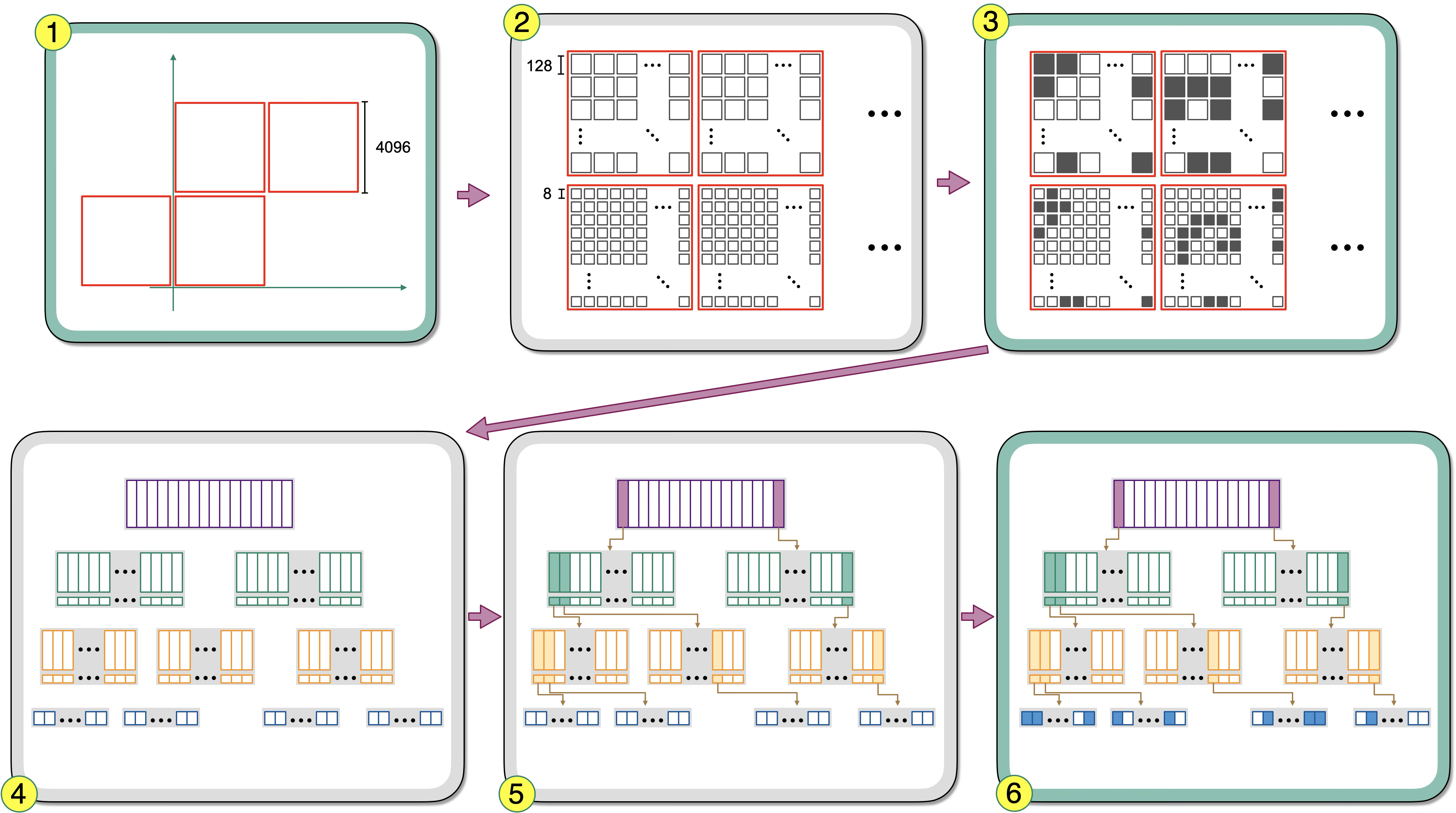}
\caption{\textbf{Process pipeline of topological operators.} For each step we indicate whether it is operation-specific (outlined in green) or generic (outlined in grey) across all operations. [\hyperref[par:step1-root-tile-preprocessing]{Step 1}; op-specific] A candidate set of root tiles (each spanning $4096^3$ voxels) is introduced to conservatively cover the resulting topology. [\hyperref[par:step2-hierarchical-mask-marking]{Step 2}; generic] For each candidate root tile, we allocate a dense bit map (representing the upper tree node) encoding the presence of each lower tree node (spanning $128^3$ voxels each; top row) and each leaf node ($8^3$ voxels each; bottom row). [\hyperref[par:step3-parallel-scratchpads]{Step 3}; op-specific] Depending on the operation, we mark in parallel all lower/leaf nodes that must be materialized, using the previously allocated bitmap scratchpads. [\hyperref[par:step4-allocation]{Step 4}; generic] We enumerate all necessary tree nodes (pruning empty ones) and monolithically allocate storage for the resulting Index-Grid. [\hyperref[par:step5-structural-reconnect]{Step 5}; generic] The connectivity of the tree is rebuilt in parallel from the lower/leaf bitmap scratchpads. [\hyperref[par:step6-voxel-population]{Step 6}; op-specific] The previous stage leaves a properly allocated sparse bitmap that tightly covers the active voxel set, but with no voxels yet marked active; we populate the active voxel flags in an operation-dependent way to complete the operation. [\hyperref[par:step7-post-processing]{Step 7}; generic (not pictured)] A final bottom-up sweep updates any remaining metadata (voxel offsets, bounding boxes, checksum etc).
}
\label{fig:pipeline}
\end{figure*}

Figure \ref{fig:pipeline} illustrates the conceptual pipeline that our topological operators follow. Several elements of this pipeline are common to all operators, while others are specialized per operation; we call these out explicitly in the description of \cref{fig:pipeline}.

\subsection{Topology Enumeration and Allocation}\label{sec:topology-enumeration}

Index-Grid reconstruction follows a multi-phase pipeline designed to saturate GPU parallel resources while maintaining a linearized memory layout. It begins with \emph{topology enumeration}, where we determine the hierarchical composition (node count per tree level) of the new grid before any final memory is allocated.

\paragraph{Root Tile Table Pre-processing}\label{par:step1-root-tile-preprocessing}
We first construct a \emph{conservative envelope} of candidate root tiles for the target operation (\cref{fig:pipeline}; Step 1). While a tight estimate is ideal, we prefer to over-provision and speculatively create tiles that may go unused, as these are removed in subsequent steps. The construction of this candidate set is operator-specific: for \emph{merging}, we take the union of all input root tiles; for \emph{pruning}, we retain the existing set (deferring removal of unused ones). For \emph{dilation}, \emph{coarsening}, and \emph{refinement}, we analyze the index-space bounding box of each existing tile and instantiate candidate neighbors wherever the operation may trigger spatial expansion. These operations run on the CPU (e.g., via a hash table), as the  root table is a minuscule fraction of the tree's total memory footprint.

\begin{figure}[t]
\centering
\includegraphics[width=\columnwidth]{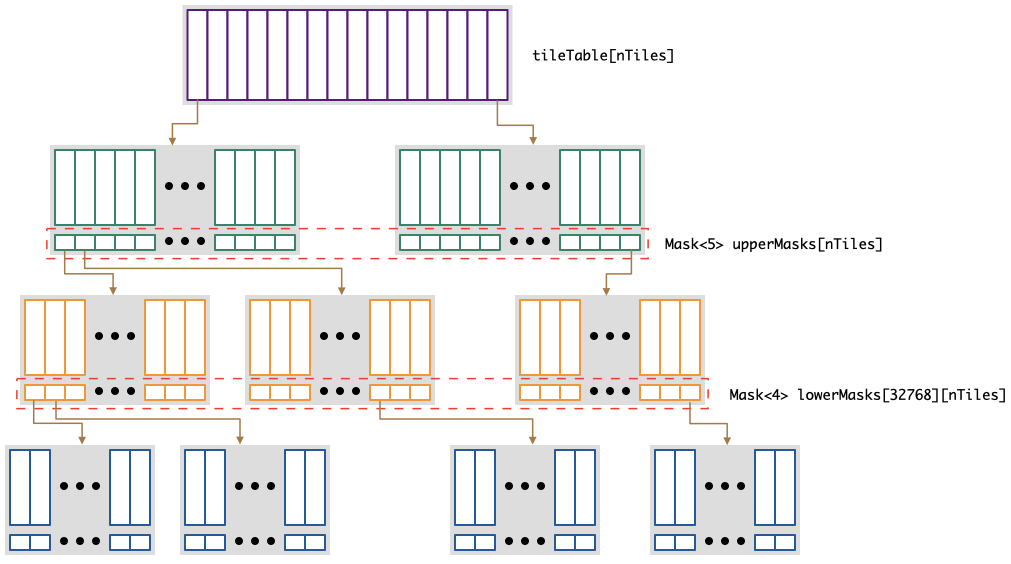}
\caption{\textbf{Topology Enumeration via Linearized Scratchpads.} To facilitate parallel reconstruction, we map the tree onto a \emph{conceptually dense} intermediate representation of serialized bit-mask arrays. Each candidate root tile is associated with one \texttt{Mask<5>} in the \texttt{upperMasks} array and a block of $32^3$ \texttt{Mask<4>} entries in the \texttt{lowerMasks} array. This layout resolves any coordinate to a specific bit via $O(1)$ spatial-to-index mapping, capturing the exact occupancy of the next epoch before the final NanoVDB buffer is materialized.}
\label{fig:nodemasks}
\end{figure}

\paragraph{Hierarchical Mask Marking}\label{par:step2-hierarchical-mask-marking}
We next allocate two bit-encoded scratchpad structures to enumerate the result tree nodes precisely (\cref{fig:pipeline}; Step 2). To visualize their encoding, consider as in \cref{fig:nodemasks} a \emph{conceptual dense NanoVDB tree} that starts at the candidate root tiles  and branches out to a dense set of child nodes at every finer level, down to leaf nodes. In NanoVDB, each non-leaf node maintains a bitmap (\texttt{childMask}) marking all its geometric children that contain active voxels. Our scratchpads store, sequentially, all the \texttt{childMask} bitmaps at the two internal levels of the tree, as seen in \cref{fig:nodemasks}.

NanoVDB defines a templatized class \texttt{Mask<N>} for storing bitmaps of dimension $(2^N)^3$. At the upper level, these are stored in an array, \texttt{upperMasks}, of \texttt{Mask<5>} bitmaps ($32^3$ bits each). At the lower level, we allocate a 2D array, \texttt{lowerMasks}, consisting of $32^3$ \texttt{Mask<4>} entries per root tile, indexed by the parent tile ID and the local linear offset of the child node. The storage footprint is remarkably modest: 4\,KB for \texttt{upperMasks} and 16\,MB for \texttt{lowerMasks} of each candidate root tile. Since a single root tile covers $4096^3$ voxels, practical grids rarely exceed a few dozen tiles. Once a tile's occupancy exceeds a mere $0.01\%$, the footprint of a single scalar float channel already outweighs these temporary bookkeeping structures.

\label{par:step3-parallel-scratchpads}These structures serve as parallel scratchpads for capturing the target topology (\cref{fig:pipeline}; Step 3). For each operator, we evaluate the input topology in parallel to determine which leaf nodes must exist in the output epoch. For \emph{merging}, an active leaf in either input mandates inclusion; for \emph{pruning}, a leaf persists if any of its active voxels remain after mask application. (The more intricate logic of \emph{dilation} is addressed in \cref{sec:dilation}.) Necessary leaf nodes are directly marked in the scratchpad bitmaps along with the lower node parents. Because the scratchpads are dense, threads  determine the target bit location statically via a simple root-tile table probe. We use \texttt{atomicOr} reductions to populate the bitmaps in parallel, giving high throughput without complex synchronization.

\paragraph{Prefix Sums and Monolithic Allocation}
Once the scratchpad bit-masks are marked, we resolve the final hierarchical counts through a series of parallel prefix sums. We define three intermediate count arrays: \texttt{upperCounts}, \texttt{lowerCounts}, and \texttt{leafCounts}.

\begin{center}
\small
\begin{tabularx}{\columnwidth}{@{}llX@{}}
\toprule
\textbf{Array} & \textbf{Size} & \textbf{Definition (for index $i$)} \\ \midrule
\texttt{upperCounts} & $n$Tiles & $1$ if $i$-th upper node has any active\\
                     &          & lower node children; else $0$ \\ \addlinespace
\texttt{lowerCounts} & $n$Tiles $\times$ 32K & $1$ if $i$-th lower node has any active\\
                     &                       & leaf node children; else $0$ \\ \addlinespace
\texttt{leafCounts}  & $n$Tiles $\times$ 32K & Number of leaf node children of the $i$-th\\
                     &                       & lower node (\texttt{popcnt} of the lower node's \texttt{childMask}) \\ \bottomrule
\end{tabularx}
\end{center}

 \noindent For each potential lower internal node $i$ in the flattened sequence, we set \texttt{lowerCounts[i]} to $1$ if the corresponding \texttt{Mask<4>} -- which identifies the presence of its $16^3$ potential \emph{leaf} children -- is non-empty, and \texttt{leafCounts[i]} to the bit population count (\texttt{popcnt}) of that same mask. \texttt{upperCounts} is populated analogously, from the occupancy of the respective \texttt{Mask<5>} bit-masks.

\label{par:step4-allocation}To resolve global memory addresses, we compute inclusive prefix sums for these count arrays, using the NVIDIA CUB library for the parallel scans and reductions. The terminal element of each resulting offset array provides the exact global sum: the total number of nodes at that level of the hierarchy. Knowing these topological requirements \emph{a priori}, we perform a single \emph{monolithic allocation} for the entire tree structure (\cref{fig:pipeline}; Step 4). This eliminates fine-grained heap allocations and multiple synchronization points, and stores the new grid epoch in a perfectly packed, contiguous buffer.

\subsection{Index-Grid Materialization and Voxel Population}\label{sec:topology-materialization}

With the destination buffer allocated and the hierarchical offset arrays computed, we proceed to materialize the final Index-Grid in three parallel stages: structural reconnection, voxel-level bit-mask population, and topological post-processing.

\paragraph{Structural Reconnection and Root Pruning}\label{par:step5-structural-reconnect}
The first stage  ``wires'' the parent-child relationships within the new buffer (\cref{fig:pipeline}; Step 5). Each thread responsible for a node in the scratchpad uses its corresponding prefix sum from the \texttt{offsets} arrays to resolve its absolute address in the destination buffer. This phase also performs the final pruning at the root level. While the CPU pre-processing phase speculatively created a \emph{conservative envelope} of candidate root tiles, the \texttt{upperOffsets} array now acts as a contraction map: a candidate tile $i$ is written to the final \texttt{tileTable} only if its entry in the offset array indicates it is non-empty.

For every surviving tile and internal node, we update the relative child pointers by sampling the starting offset of its children's range from the next level's offset array (e.g., an internal node at index $j$ retrieves its child pointer from \texttt{leafOffsets[j]}). At the end of this pass the tree is fully valid for traversal, although the leaf-level voxel masks remain empty.

\paragraph{Operator-Specific Voxel Population}\label{par:step6-voxel-population}
With the skeletal structure established, we populate the leaf-level voxel masks (\texttt{valueMask}) in an operation-specific manner (\cref{fig:pipeline}; Step 6). Separating structural reconnection from bit-mask filling is a key optimization, as the latter can be performed in-place within the already-allocated hierarchy. For operators such as \emph{merging} and \emph{pruning}, threads perform bitwise \texttt{OR} or \texttt{AND} operations between input grids and masks, writing directly to the new leaf nodes. For \emph{coarsening} and \emph{refinement}, we perform structured 2:1 bit-permutations to downsample or upsample the occupancy patterns. The logic for morphological \emph{dilation} is more complex and is detailed in \cref{sec:dilation}.

\paragraph{Topological Post-Processing}\label{par:step7-post-processing}
The final stage of the pipeline performs global bookkeeping. We compute a final prefix sum across the newly populated leaf bit-masks to assign a unique \emph{voxel offset} to every leaf node. This count provides the total active voxel population of the tree and defines the size of the linearized data sidecar discussed in \cref{sec:vbm}. Finally, a parallel reduction updates the index-space bounding boxes and computes the grid checksum, leaving the grid  fully compliant with the NanoVDB specification.

\subsection{Morphological Operators: Bit-Parallel Dilation}\label{sec:dilation}

While merging and pruning follow a straightforward implementation path, morphological dilation presents a unique challenge: an active voxel at the boundary of a \emph{leaf node} may require materializing adjacent neighbors to accommodate the dilation stencil's footprint. For a standard $3 \times 3 \times 3$ stencil, a single boundary voxel may instigate the creation of up to seven neighboring leaf nodes (at a leaf corner), while the collective occupancy of a leaf can influence all 26 neighbors. To resolve this without expensive tree traversals for neighbor access, we implement a \emph{hierarchical bit-parallel dilation} that decouples topological expansion from voxel-level population.

\paragraph{Leaf-level Neighbor Influence Masks}

Dilation begins with a parallel sweep in which each leaf node is processed by an individual CUDA thread, determining whether the dilated footprint of the leaf's active voxels intersects any of its 26 leaf neighbors and thus necessitates their materialization in the result tree.

We recall that the $8 \times 8 \times 8$ voxel occupancy mask is bit-packed into eight \texttt{uint64\_t} variables (in \texttt{Mask<3>}).
For a standard $3 \times 3 \times 3$ box dilation, we infer whether each leaf neighbor must materialize using simple bitwise operations on these packed words: combining slices via bitwise \texttt{OR} and isolating the ``faces'' of the $8^3$ volume via bitwise \texttt{AND} with predetermined masks yields a 27-bit \emph{influence mask} per leaf, indicating whether each of its leaf neighbors needs to be introduced. These masks accommodate even wider dilation scenarios; for any stencil with a radius $R \leq 8$, the influence of an $8^3$ leaf node is strictly confined to its immediate $3^3$ cluster of leaf neighbors. This leaf-level computation is the only step that inspects the dilation stencil itself; all subsequent steps depend only on the resulting mask of affected leaf neighbors.

\subsubsection*{Leaf-Major to Tap-Major Transposition}
To avoid the high contention of irregular atomic updates, we transpose these influence masks in parallel in shared memory, shifting from a \emph{leaf-major} ``scatter'' (each leaf updates 27 taps) to a \emph{tap-major} ``gather'' (the lower node computes 27 uniform spatial shifts). The left of the following illustration shows the state prior to this operation: for each of the $16^3$ leaf node children of a \emph{lower} internal node, we have a 27-tap bit stencil indicating \emph{based on the voxels within the leaf node that this stencil is centered on, which of the $3\times 3\times 3$ neighboring leaf nodes should be made active?}

\begin{figure}[h]
\centering
\includegraphics[width=\columnwidth]{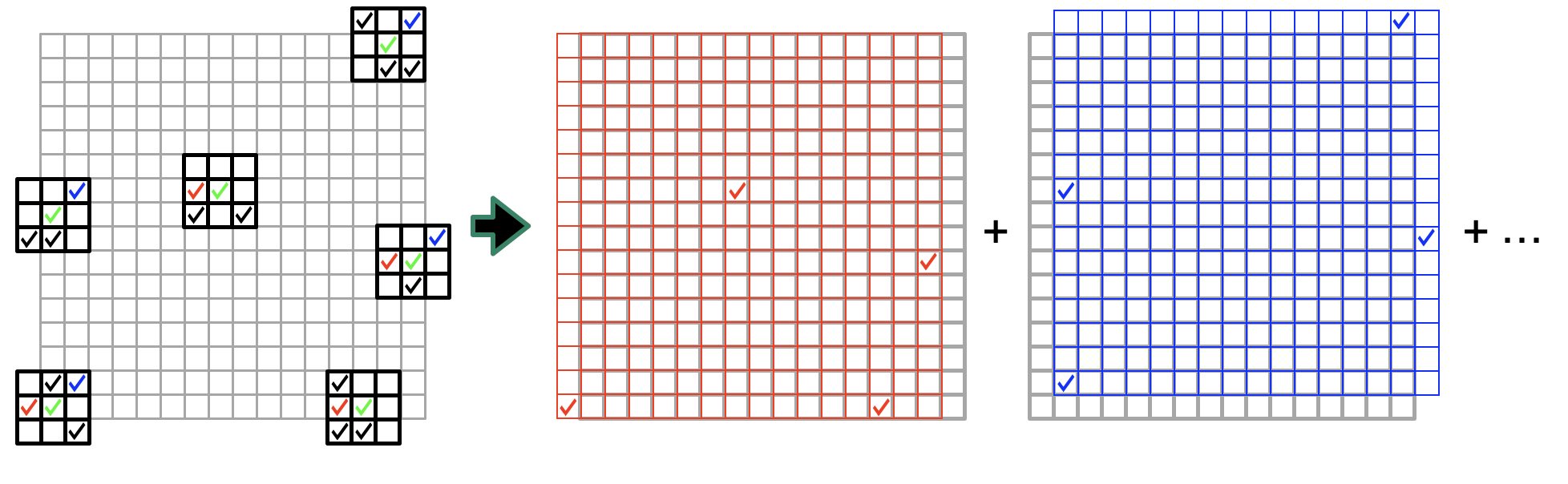}
\caption{\textbf{Leaf-major to tap-major transposition.} Per-leaf neighbor-influence stencils (left) are reorganized into one bitmask per stencil offset (right), enabling warp-collective processing.}
\end{figure}

\noindent We transpose this (leaf-major) grid-of-stencils into a tap-major stencil-of-grids (each grid encoded as a NanoVDB \texttt{Mask<4>}), as seen on the right side of the illustration above. We perform this transposition at warp level, using \texttt{CUB::WarpReduce} to aggregate influence bits across lanes, tallying the ``votes'' of 32 leaves into a single bit-parallel write. The resulting \texttt{offsetMasks} represent the collective topological footprint of the lower internal node for every neighbor offset.

\subsubsection*{Warp-Collective Stencil Convolution}
We finally combine the 27 ``offset'' \texttt{Mask<4>} bitmaps to compute the leaf nodes that should be marked active, both within the extent of this lower node and its $3\times 3\times 3$ neighbors. As a building block we implement a warp-level operator \texttt{MaskShift<di,dj,dk>}, a 3-dimensional bit shift on a \texttt{Mask<4>} with compile-time signed shift distances; this uses warp-level shuffle and rotation instructions to exploit bit-level parallelism. We accumulate the result by shifting each \texttt{offsetMask} by its own tap offset and combining the shifted masks with a bitwise \texttt{OR}: a bit set at leaf slot $p$ of tap $(di,dj,dk)$ indicates that slot $p+(di,dj,dk)$ must be materialized. Bits displaced past the $16^3$ extent of the lower node are accumulated into the mask of the corresponding neighboring lower node, so the kernel yields 27 \texttt{resultMasks}: one for this lower node and one for each of its 26 neighbors.

\subsubsection*{Global Atomic Merging}
The kernel concludes by committing the 27 resulting neighbor masks to the global scratchpads defined in \cref{sec:topology-enumeration}. For any neighbor mask containing active bits, the thread probes the \texttt{dilatedRoot} to find the target tile and uses \texttt{atomicOr} to update \texttt{upperMasks} and \texttt{lowerMasks}. This ``blackboard'' approach lets thousands of threads concurrently contribute to the new topology, which is then finalized via the prefix-sum pipeline.

\subsubsection*{Voxel Mask Reconstruction}
One further dilation-specific computation is required. At Step 6 of \cref{fig:pipeline} the tree topology holding the dilation result exists, but no voxels have been set active. We perform this as a bit-parallel \emph{gather}: every leaf node composes its \texttt{valueMask} (a bit-encoded \texttt{Mask<3>}) by fetching the \texttt{valueMask}s of neighboring leaf nodes from the source topology and combining them with bit-parallel shifts.

\subsection{Sidecar Update Operators (Inject)}\label{sec:sidecar-ops}

The operations described thus far affect only the \emph{topology} (active voxel set) of a sparse Index-Grid, leaving the associated data payload (the ``sidecar'') unaffected. We deliberately keep the sidecar update as a separate operator, since with our GPU-optimized implementation the topological updates typically complete in less time than it takes to stream the data payload through memory. We introduce the following abstraction:

\begin{myshaded}
\begin{verbatim}
injectData(srcGrid, dstGrid, srcSidecar, dstSidecar)
\end{verbatim}
\end{myshaded}

\noindent This operation takes the source and target  topologies (\texttt{srcGrid},\allowbreak{} \texttt{dstGrid}) and their respective sidecars. It operates in accordance with the following semantics: any voxel that is present in the \emph{intersection} of the two Index-Grids triggers a copy of the data from the location in the source sidecar to the  location the same coordinate maps to in the destination sidecar; all other data in the destination sidecar is left untouched. This definition has proven fully adequate for all practical applications of the operators in this section.

Our implementation maximizes memory coalescing, especially where one topology is a proper subset of the other (in which case either reads or writes are coalesced). We assign each pair of source\slash destination leaf nodes with the same origin to an individual warp, leveraging the guarantee that data for active voxels within a single leaf node is sequential in the sidecar. We also provide variants of \texttt{inject} for strided sidecars and for sidecars holding multi-dimensional (vector-valued) features, as is common in fVDB machine learning workloads.
\section{Voxel-sequential access and the VoxelBlockManager}\label{sec:vbm}

A fundamental feature of Index-Grids \cite{Williams:2024:fVDB} is the compactness of storage they enable: only values associated with \emph{active voxels} need to be stored in the auxiliary sidecar, in contrast to prior iterations of NanoVDB, which quantize data allocation upwards to full $8\times 8\times 8$ blocks corresponding to the domain of each leaf node. It is natural to seek a matching compactness of computation, especially for streaming and stencil-oriented workloads. Ideally, a data-parallel operation such as applying a finite difference stencil to all active voxels would iterate over the linear sequence of  active voxels, regardless of how they are distributed across leaf nodes or how sparsely occupied those leaf nodes are in different parts of the domain.

Implementing such an iteration paradigm, which we call \emph{voxel-sequential access}, is significantly more challenging on the GPU. On the CPU, the typical parallelization pattern in both OpenVDB and NanoVDB distributes leaf nodes across threads and processes their active voxels in a single-threaded iteration. On the GPU, the canonical pattern launches kernels over leaf nodes, with an entire thread block  to process all $8^3$ voxels of a leaf in parallel, active or not. The load balance difficulty is clear: any warp mapped to a 32-tuple of voxels that are not all active wastes SIMT throughput on the inactive fraction. The same argument applies to the CPU; the loss is simply more evident on the GPU, where SIMT computation is the default, whereas explicit  SIMD on the CPU  is typically an intentional, targeted optimization.

{}
We introduce \emph{voxel-sequential access}, a linear iteration paradigm over only active voxels, in the order that Index-Grid defines.
An Index-Grid is functionally a \emph{forward map} from a Cartesian voxel coordinate $(i,j,k)$ to its sequential active-voxel index; voxel-sequential access requires the \emph{inverse} of this map, which VDB has not previously provided. We introduce a new NanoVDB acceleration structure, the \emph{Voxel Block Manager} (VBM) to encode this inverse map: given the sequential index of an active voxel, it recovers the voxel's position in the tree---its containing leaf and in-leaf offset---and by extension its Cartesian coordinate. This section defines the structure, highlights its very low storage footprint, and illustrates its use in  voxel-sequential  streaming and stencil operations.

\subsection{VoxelBlockManager metadata and initialization}\label{sec:vbm-generation}

\begin{figure}[t]
\centering
    \includegraphics[width=\columnwidth]{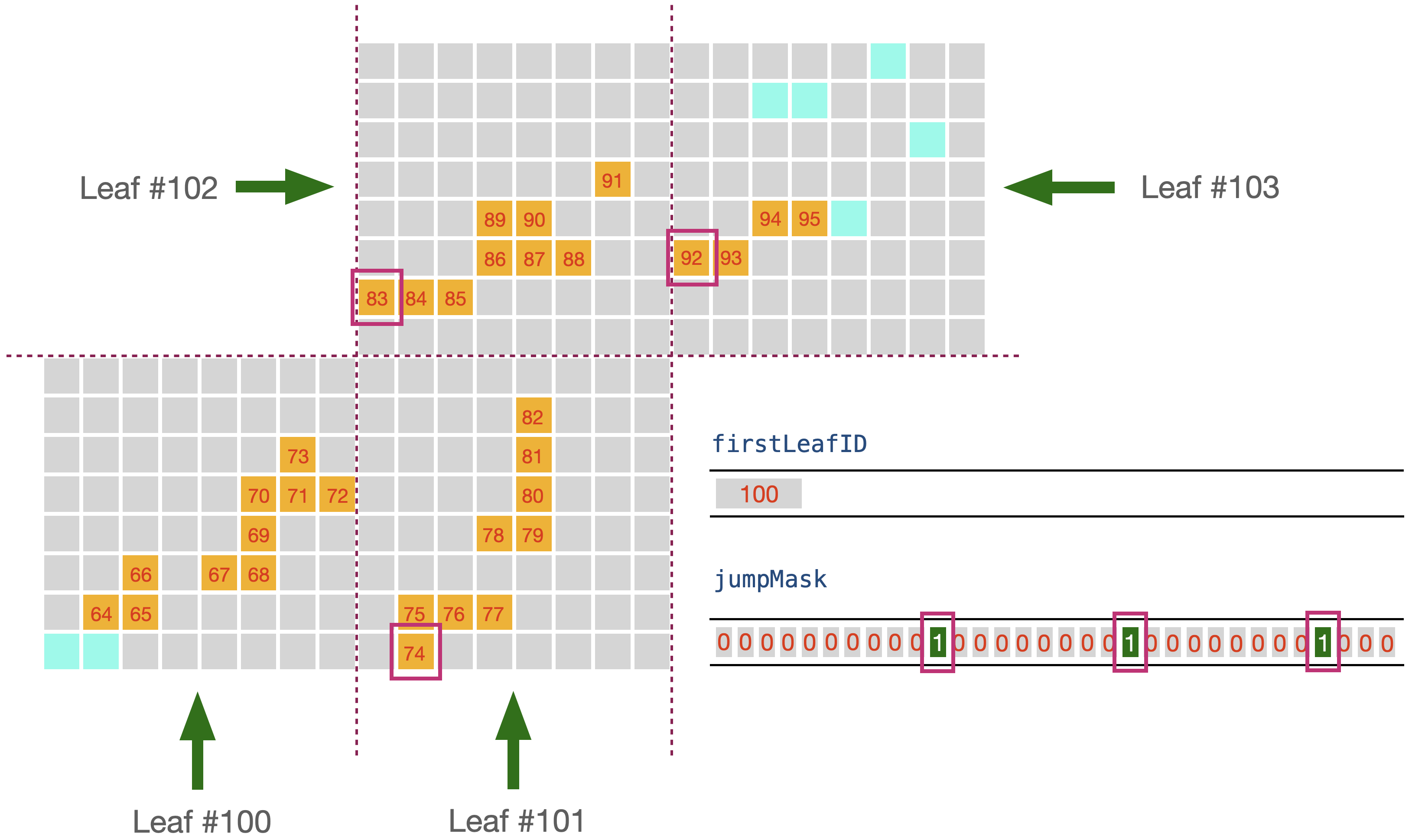}

\caption{\textbf{VBM metadata for a $W{=}32$ block.} The 32 active voxels at sequential indices $64\ldots 95$ (orange) span four leaf nodes of the underlying Index-Grid; a few active voxels with indices outside this block appear in cyan. The metadata for the block reduces to a single $\texttt{firstLeafID}=100$ (the leaf containing the first voxel in the block) plus a 32-bit $\texttt{jumpMap}$ whose set bits at positions $10$, $19$, and $28$ mark the intra-block offsets at which each subsequent leaf begins.}
    \label{fig:vbm-metadata}
\end{figure}

The VBM is parameterized by a compile-time
\emph{block width} $W \ge 64$, taken to be a multiple of $64$. Conceptually, $W$ is the width of the voxel interval we operate on in a data-parallel fashion. On the GPU it is typically chosen to match the thread block size of the operation being implemented ($W=128$ or $W=256$ in most of our examples). We partition the sequence of active voxels into aligned intervals of size $W$, which we call VBM \emph{blocks}. For each block, the VBM stores two metadata items (Figure \ref{fig:vbm-metadata}):

\begin{itemize}
    \item \texttt{firstLeafID}: a \texttt{uint32\_t} per block, storing the index of the leaf node containing the block's first active voxel.
    \item \texttt{jumpMap}: one bit per active voxel, packed into \texttt{uint64\_t} words ($W/64$ per block). Bit $p$ ($p>0$) is set if the active voxel at intra-block position $p$ is the \emph{first} active voxel of a new leaf node.
\end{itemize}

\noindent Together these consume at most $1.5$ bits per active voxel, independent of grid size, payload type, or leaf occupancy.
NanoVDB Index-Grids guarantee that the active voxels within a single leaf node occupy a contiguous index interval. The semantics of the VBM implicitly assume two further properties:
    \begin{itemize}
    \item No leaf nodes have zero active voxels (such a leaf node would be functionally irrelevant to the coordinate-to-index map the grid defines).
    \item {}Active voxel indexing is contiguous across leaf nodes: if $K$ is the index of the last active voxel in leaf node $L$, the first active voxel in leaf node $L+1$ has index $K+1$.
    \end{itemize}
    \noindent Although neither property is stipulated by the core NanoVDB data structure, both are de facto enforced by all current NanoVDB tools that generate Index-Grids. Given their utility in enabling the VBM, we mandate them as invariants in applications that use it.

Finally, the \texttt{firstLeafID} and \texttt{jumpMap} arrays are computed in parallel. For each leaf we identify all overlapping \emph{blocks}. Blocks overlapping a strict prefix of the leaf's voxel interval have the leaf's starting position marked in their \texttt{jumpMap} via \texttt{atomicOr}; blocks beginning inside the leaf have their \texttt{firstLeafID} set to its index. Construction cost is typically sub-millisecond for Index-Grids with up to $10^7$ active voxels.

\subsection{VBM Decoding in stencil-based kernels}\label{sec:vbm-consumption}

A kernel using the VBM is launched with one $W$-thread CUDA block per VBM block. Every such kernel begins by decoding the VBM metadata at thread block granularity, consuming the block's \texttt{firstLeafID} entry and \texttt{jumpMap} row and writing two arrays to shared memory:

\begin{itemize}
    \item \texttt{leafIndex[$W$]}: the index of the leaf containing each thread's active voxel.
    \item \texttt{voxelOffset[$W$]}: the local (Cartesian) offset of that voxel within its leaf, encoded as a \texttt{uint16\_t} in $xyz$-octal form (three bits per axis).
\end{itemize}

A section of the decoded arrays, corresponding to Leaf $\#101$ from Figure \ref{fig:vbm-metadata} is illustrated below:

\begin{figure}[h]
\centering
\includegraphics[width=\columnwidth]{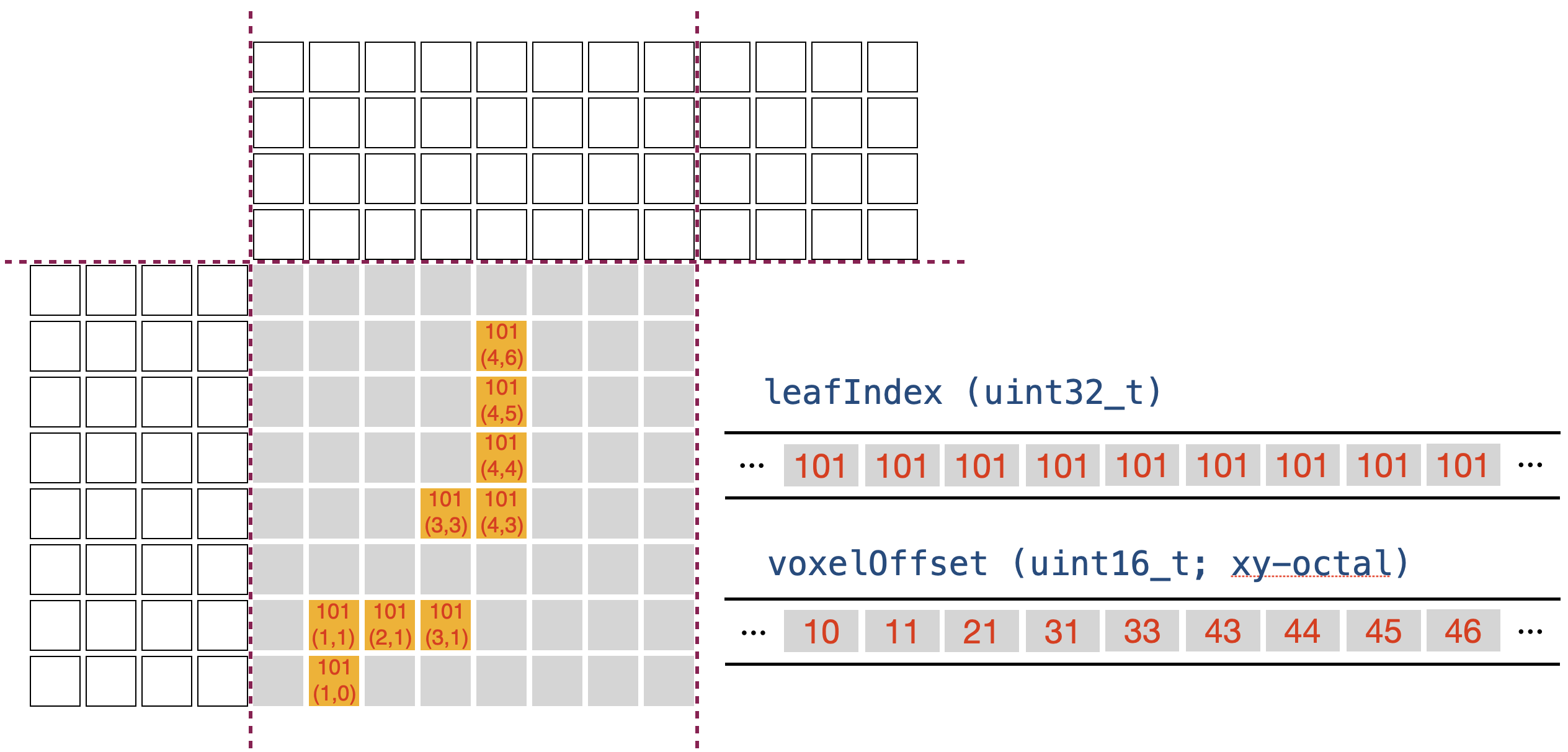}
\end{figure}

Decoding operates as follows. The number of leaf nodes overlapping the current block is the bit population count of the \texttt{jumpMap}, plus one. For each such leaf  we visit all 512 ($8^3$) voxels in parallel and probe whether each is active and, if so, what its linear index is -- both fundamental NanoVDB operations, accelerated via built-in \texttt{popcnt} instructions on the GPU. Active voxels  update the corresponding entries of \texttt{leafIndex} and \texttt{voxelOffset}  in shared memory. The operation requires no delicate synchronization and incurs no global memory traffic, so for all but the most extreme  leaf-level sparsity ($>95\%$ inactive voxels per leaf node), its cost is substantially lower than that of accessing the sidecar values themselves.

\subsubsection*{Stencil-tap Optimization} For stencil applications we  employ a further optimization that mitigates the cost of computing  linear indices for neighboring voxels. For common, narrowly-supported stencils, the number of leaf nodes  a stencil can touch is substantially lower than its number of taps. The fifth-order Hamilton-Jacobi WENO stencil used in our fluid simulation examples (\cref{sec:fluid_sim}) has 19 taps, at grid offsets $(-3\ldots 3,0,0)$, $(0,-3\ldots 3,0)$, and $(0,0,-3\ldots 3)$, yet can touch at most \emph{four} leaf nodes (the one containing the stencil center, and at most three more). We identify the required neighbors and fetch them in bulk via NanoVDB's \texttt{probeLeaf}; the indices of all 19 stencil taps are then computed with fully warp-synchronous, non-divergent operations. Similar optimizations apply to other stencil types; for example a $3\times 3\times 3$ convolution stencil for machine learning workloads touches at most 7 neighboring leaf nodes (typically far fewer), to compute all 27 taps.

\section{Applications}\label{sec:applications}

The dynamic topology operators (\cref{sec:topology}) and the voxel-sequential access paradigm (VBM, \cref{sec:vbm}) are general-purpose primitives that compose into algorithms spanning a wide variety of real-world domains. We demonstrate this with applications from 3D perception and reality capture (\cref{sec:reality_capture}), fluid and gas simulation (\cref{sec:fluid_sim}), and large-scale shape voxelization and rendering (\cref{sec:large_udf}).  Furthermore, the algorithms built from these compositions are often more memory-efficient and performant than their counterparts in the literature. The details of the hardware used for our experiments can be found in Appendix \ref{appendix:hardware-config}.

\subsection{Perception and Reality Capture}\label{sec:reality_capture}

Together, the topology operators of \cref{sec:topology} and the voxel-sequential paradigm of \cref{sec:vbm} undergird an implementation of established reality-capture techniques: truncated signed distance fields (TSDF) for surface reconstruction~\cite{curless1996tsdf, newcombe2011kinectfusion}, log-odds occupancy for path planning~\cite{hornung2013octomap}, the Euclidean signed distance field (ESDF)
for collision checking~\cite{oleynikova2017voxblox, han2019fiesta}, and dynamic-scene decay to filter moving objects. Each operation is an integrator expressed in a few hundred lines of C++/CUDA, with per-voxel
state held in PyTorch tensors aligned to the linear order of each grid's active voxels. We present TSDF fusion as a representative composition and cover the rest in Appendix \ref{appendix:rc-full}.

\begin{figure*}[t]
\includegraphics[width=\textwidth]{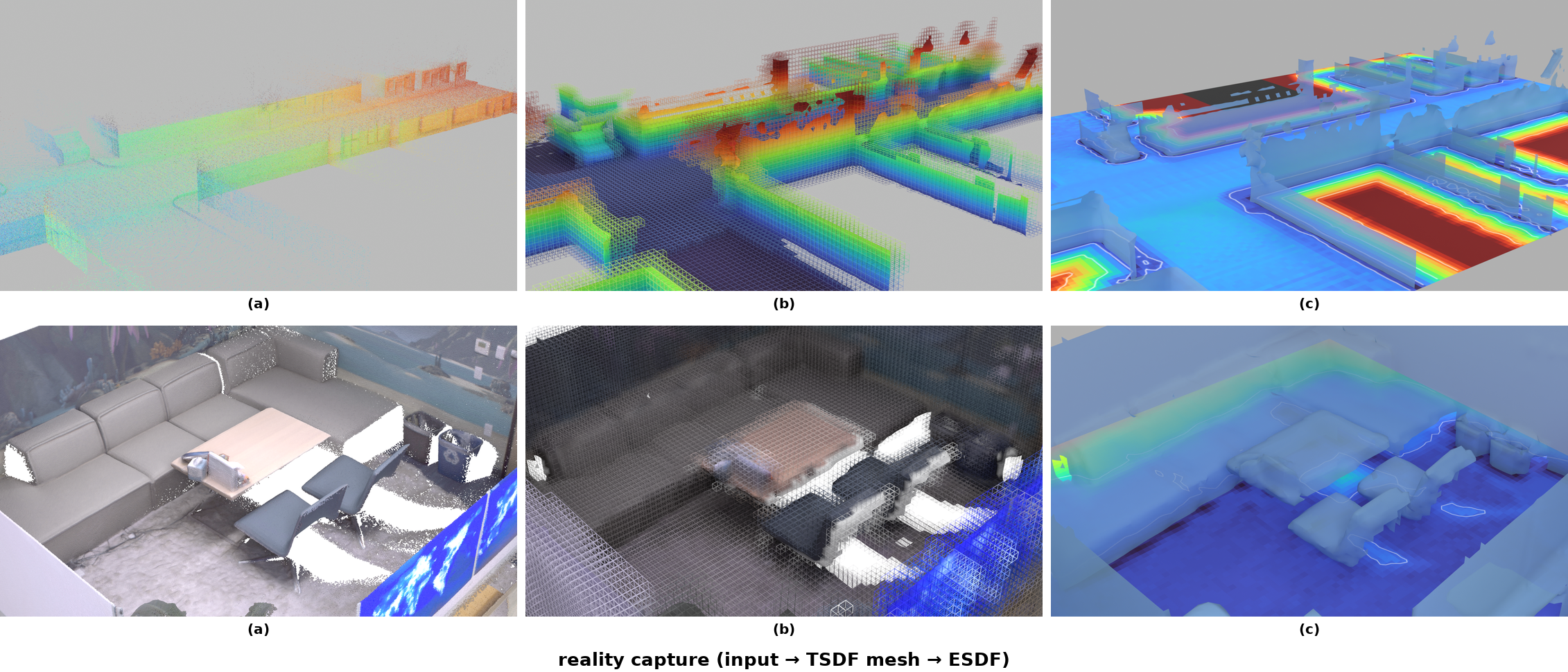}
\caption{Starting from an input LiDAR sweep or a sequence of depth images, we fuse a TSDF surface mesh and subsequently derive a Euclidean signed distance field (ESDF) to facilitate collision checking and path planning.
    Mai City synthetic LiDAR (top row, voxel size $0.33$~m) and
    Replica \textsc{office0} RGB-D (bottom row, voxel size $0.05$~m),
    with (a) input sensor sweeps coloured by time (Mai City) and by
    source-pixel RGB (Replica), (b) fused TSDF mesh with the
    underlying voxel grid drawn as a faint wireframe scaffold, and
    (c) the mesh rendered translucently over a horizontal slice of
    the derived ESDF (turbo colormap, red~=~far from any obstacle).}
\label{fig:reality-capture-pipeline}
\end{figure*}

\paragraph{TSDF integration as a representative composition.}
We integrate a depth or LiDAR observation $Z$ at time $t$ into a grid $G_{t-1}$ with per-voxel weights $w_{t-1}$ and signed distances $d_{t-1}$ using \texttt{voxelsToGrid}, \texttt{mergeGrids}, and \texttt{inject} from \cref{sec:topology} together with the per-ray HDDA traversal~\cite{10.1145/2614106.2614136,Williams:2024:fVDB}. The depth case carries a camera projection $\Pi$; the LiDAR case carries a sensor origin $o$. A voxel handle $v$ has world position $x_v = \textsc{worldCoord}(v)$. The truncated-SDF helper $\textsc{truncSDF}(d,\tau)$ returns $d/\tau$ when $|d|\le\tau$, $+1$ when $d > \tau$ (free space ahead of the surface), and the sentinel \emph{skip} when $d < -\tau$ or the underlying sample is undefined (e.g.\ the voxel projects outside the depth frame), in which case the voxel keeps its injected old value. The per-observation weight $w_{\mathrm{obs}}$ defaults to $1$ (the depth integrator optionally accepts a per-pixel weight image).

\Cref{alg:tsdf} shows the resulting integration. Lines~1--3 are shared across modalities: they build a shell of voxels within $\tau$ of the new observation, union it into a fresh grid $G_t$, and inject the old sidecars onto $G_t$, zeroing the newly created voxels. The update phase differs by modality. \emph{Depth} dispatches one thread per voxel of $\mathit{shell}$ and writes $(d_t[v], w_t[v])$ in place with a weighted running average---no atomics because each voxel has a single writer. \emph{LiDAR} dispatches one thread per input point, walks the active voxels along the sensor ray $\overline{op}$ via HDDA, and accumulates into the running-sum sidecars $(s_d, s_w) = (d_t\cdot w_t,\, w_t)$ via \texttt{atomicAdd}; a final $d_t \leftarrow s_d/s_w$ recovers the running average. The same scaffold powers occupancy mapping (with a Bayesian log-odds add as the per-voxel update) and the dilated-grid stage of ESDF; dynamic-scene decay is \texttt{pruneGrid} plus \texttt{inject} with no integrator-side code at all (Appendix \ref{appendix:rc-full}).

\begin{algorithm}[h]
\DontPrintSemicolon
\KwIn{Persistent $(G_{t-1}, d_{t-1}, w_{t-1})$; observation $Z$ with projection $\Pi$ (depth) \emph{or} sensor origin $o$ (LiDAR); truncation $\tau$; per-observation weight $w_{\mathrm{obs}}$.}
\KwOut{$(G_t, d_t, w_t)$.}
\tcc{Topology grow + inject (shared across modalities)}
$\mathit{shell} \leftarrow \textsc{voxelsToGrid}\!\left(\textsc{worldPoints}(Z) \oplus \textsc{ball}(\tau)\right)$\;
$G_t \leftarrow \textsc{mergeGrids}(G_{t-1}, \mathit{shell})$\;
$(d_t, w_t) \leftarrow \textsc{inject}\!\left(G_{t-1}\!\to\!G_t,\,(d_{t-1}, w_{t-1}),\,\mathit{init}{=}0\right)$\;
\BlankLine
\uIf(\tcp*[h]{per-voxel dispatch; no atomics}){$Z$ \textnormal{is a depth image}}{
    \ForEach{$v \in \mathit{shell}$ \textnormal{in parallel}}{
        $\mathit{sdf}_v \leftarrow \textsc{truncSDF}\!\left(Z[\Pi(x_v)] - z_{\mathrm{cam}}(x_v),\,\tau\right)$\;
        \lIf{$\mathit{sdf}_v \neq \mathrm{skip}$}{$\bigl(d_t[v], w_t[v]\bigr) \leftarrow \textsc{weightedAvg}\!\left(d_t[v], w_t[v], \mathit{sdf}_v, w_{\mathrm{obs}}\right)$}
    }
}
\Else(\tcp*[h]{per-ray HDDA; atomic running-sums}){
    $(s_d, s_w) \leftarrow (d_t\!\cdot\!w_t,\, w_t)$\tcp*{transient accumulators}
    \ForEach{$p \in Z$ \textnormal{in parallel}}{
        \ForEach{$v$ \textnormal{on} $\textsc{HDDA}(\overline{op},\,G_t)$}{
            $\mathit{sdf}_v \leftarrow \textsc{truncSDF}\!\left(\|p\!-\!o\| - \|x_v\!-\!o\|,\,\tau\right)$\;
            \lIf{$\mathit{sdf}_v \neq \mathrm{skip}$}{$\textsc{atomicAdd}(s_d[v],\,w_{\mathrm{obs}}\,\mathit{sdf}_v),\;\textsc{atomicAdd}(s_w[v],\,w_{\mathrm{obs}})$}
        }
    }
    $(d_t, w_t) \leftarrow (s_d / s_w,\,s_w)$\tcp*{normalize sums to running average}
}
\Return $(G_t, d_t, w_t)$\;
\caption{TSDF integration of a single observation. Lines~1--3 are the shared topology+inject scaffold (\texttt{voxelsToGrid}, \texttt{mergeGrids}, \texttt{inject}). \textsc{worldPoints}$(Z)$ is depth-back-projection through $\Pi$ or the identity for LiDAR. The update phase picks an iteration pattern to match the data flow: depth iterates voxels (in-place avg), LiDAR iterates rays (atomic running-sum then normalize). Occupancy reuses lines~1--3 with a Bayesian log-odds per-voxel update; ESDF reuses the scaffold with \texttt{dilateGrid} for \texttt{voxelsToGrid} (Appendix \ref{appendix:rc-full}).}
\label{alg:tsdf}
\end{algorithm}

\paragraph{Datasets and Results}
We benchmark against nvblox~\cite{millane2024nvblox}, Open3D~\cite{zhou2018open3d}, and VDBFusion~\cite{vizzo2022vdbfusion} on four datasets: Mai~City~\cite{vizzo2021poisson} synthetic Velodyne LiDAR (700-frame trajectory), Replica~\cite{straub2019replica} depth renderings produced by NICE-SLAM~\cite{zhu2022niceslam} (8-scene, 2000 frames each), KITTI Odometry~\cite{geiger2012kitti, geiger2013kittiijrr} real-sensor LiDAR (sequences 00, 02, 05; 2.2--5.1\,km), and 7-Scenes~\cite{shotton2013scenes} RGB-D for long-trajectory persistence. VDBFusion comparisons run CPU-only (OpenVDB) and are provided for completeness and qualitative analysis. All measurements use  truncation $\tau{=}3\times$ voxel size and ESDF max-distance $R{=}10\times$ voxel size throughout; nvblox's depth and LiDAR integration distances are configured to match ours rather than the upstream defaults. \Cref{tab:scale-ceiling} reports the finest voxel each system completes; per-voxel-size timings, the full nvblox-config disclosure, and qualitative mesh comparisons (\cref{fig:rc-comparisons}) are in Appendix \ref{appendix:rc-full}.

\begin{itemize}
\item \emph{Scale.} At equal memory budget, our implementations reach voxel resolutions 1.5--5$\times$ finer than the block-hash baselines, with active voxel counts up to two orders of magnitude larger. The strongest single point is \emph{1.67\,$\times$\,10\textsuperscript{9} ESDF voxels in 38.6\,GB on KITTI seq~00 at 2\,cm}, where nvblox runs out of memory.
\item \emph{Speed.} On occupancy across the 20--1.5\,cm sweep we are 3.5--11.5$\times$ faster than nvblox; on LiDAR TSDF we are 3.6--5.7$\times$ faster on Mai City and 4.8--5.3$\times$ faster on KITTI at 40\,cm, narrowing to 2.6--3.5$\times$ at 20\,cm before nvblox exhausts memory at 10\,cm. Open3D leads depth-TSDF throughput on Replica by 11--46$\times$ at 20--5\,mm before failing at 4\,mm (see Appendix \ref{appendix:rc-full}).
\item \emph{Composability.} Every integrator decomposes into the same three kinds of step: dynamic topology operators, custom CUDA kernels (per-leaf-slot, per-ray HDDA, or VBM), and PyTorch operations on sidecar tensors. Dynamic-scene decay is the limit case -- two operators, no integrator code -- with a measured \emph{78\% reduction in persistent voxels} on KITTI seq 00 over 500 frames, as moving traffic stops accumulating in the persistent grid.
\end{itemize}

\begin{table}[h]
\centering
\caption{\textbf{Voxel-size scale ceiling on a single 48\,GB RTX~6000 Ada GPU.} Finest voxel each system completes; ours reaches the listed active voxel count and peak GPU memory at its finest voxel. Open3D has no LiDAR TSDF and no ESDF integrator. $^\dagger$Open3D succeeds on 6/8 Replica scenes at 5\,mm and crashes (\texttt{illegal memory access}) at 3\,mm on all eight; we list 5\,mm as its finest universal ceiling. Per-voxel-size timing tables and the full nvblox-config are in Appendix \ref{appendix:rc-full}.}
\label{tab:scale-ceiling}
\footnotesize
\setlength{\tabcolsep}{4pt}
\begin{tabular}{@{}lcccrr@{}}
\toprule
Workload                         & nvblox   & Open3D            & Ours            & Voxels    & GB   \\
\midrule
Mai City 700\,f LiDAR TSDF       & 10\,cm   & n/a               & \textbf{1.25\,cm} & 938.7\,M & 37.1 \\
Mai City 700\,f TSDF$\,\to\,$ESDF & 10\,cm  & n/a               & \textbf{3\,cm}    & 1.28\,B  & 30.6 \\
Replica 8$\times$2000\,f TSDF    & 10\,mm   & 5\,mm$^\dagger$   & \textbf{3\,mm}   & 109\,M    & 10.3 \\
Replica 8$\times$2000\,f ESDF    & 10\,mm   & n/a               & \textbf{3\,mm}   & 438\,M    & 10.3 \\
KITTI seq 02 4661\,f TSDF        & 20\,cm   & n/a               & \textbf{10\,cm}  & 427\,M    & 19.8 \\
KITTI seq 00 100\,f ESDF         & 5\,cm    & n/a               & \textbf{2\,cm}   & 1.67\,B   & 38.6 \\
KITTI seq 05 100\,f Occupancy    & 1.5\,cm  & n/a               & \textbf{5\,mm}   & 2.35\,B   & 44.1 \\
\bottomrule
\end{tabular}
\end{table}

\subsection{Fluid and Gas Simulation}\label{sec:fluid_sim}

\begin{figure}[t]
\centering
\includegraphics[width=1\linewidth]{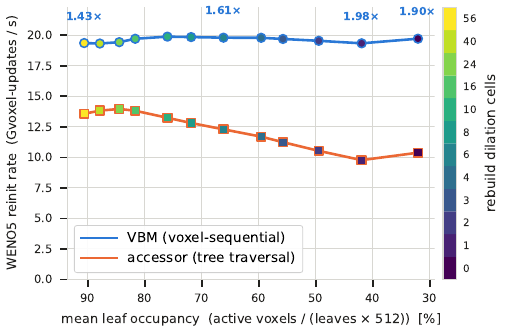}
\caption{WENO5 narrow-band reinitialization throughput for two dispatch policies—VBM (voxel-sequential, one thread per active voxel) and the accessor baseline (tree traversal, one thread per leaf slot)—against mean leaf occupancy $\equiv \texttt{activeVoxels} / (\texttt{nLeaves} \cdot 512)$ within active 8³ leaf tiles, decreasing left to right. Timing is per-call time of one inner WENO5 iteration. VBM holds near 20 Gvoxel-updates/s, its launch sized by the active set alone, while the accessor falls from 14.0 to 9.8 as more threads land on inactive slots; VBM speedup reaches 1.98×, rising as occupancy drops. Marker color denotes rebuild dilation width. Hardware details in Appendix \ref{appendix:hardware-config}.}
\label{fig:weno_sparsity}
\end{figure}

\begin{figure}[h]
    \centering
    \includegraphics[height=.9in]{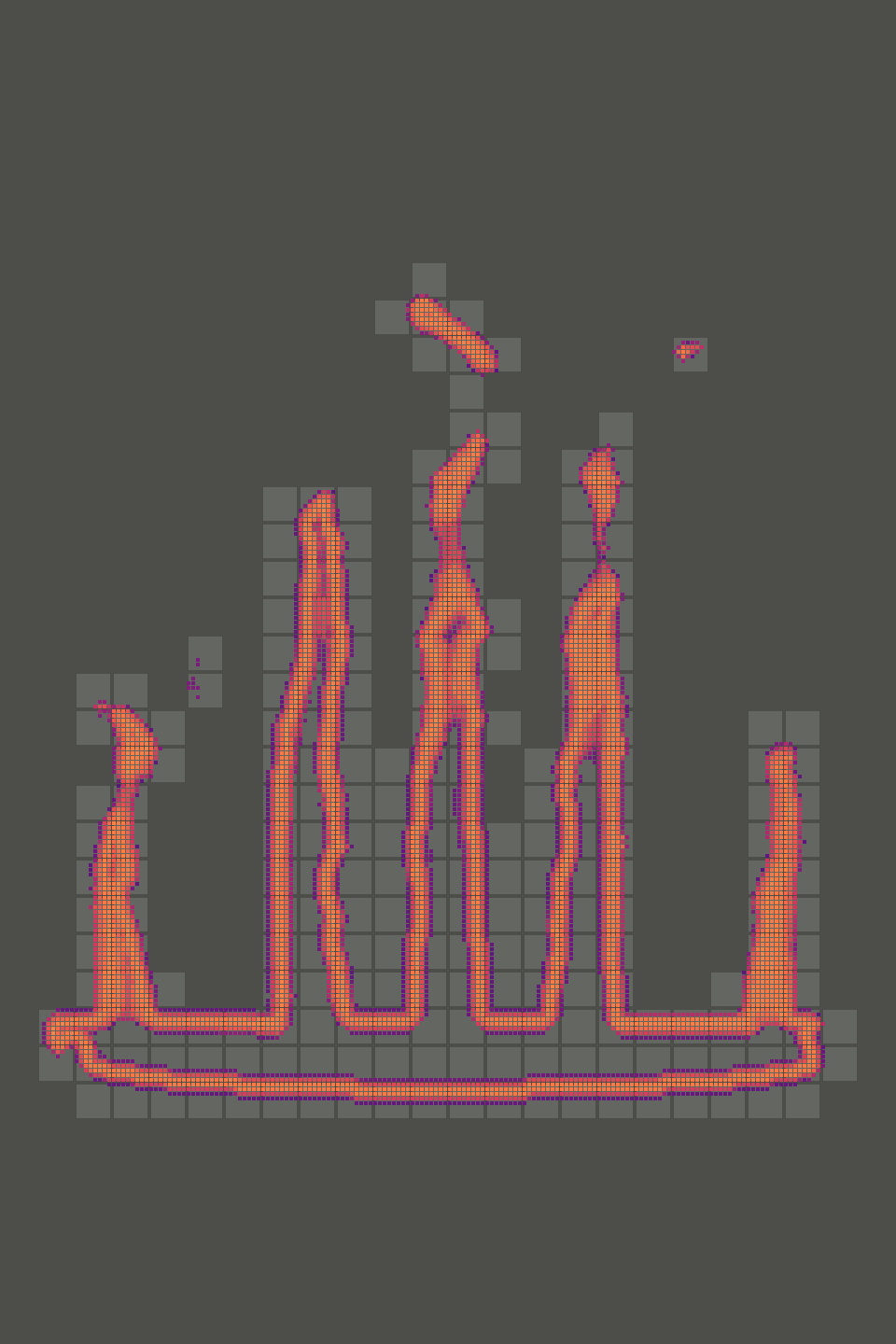}
\includegraphics[height=.9in]{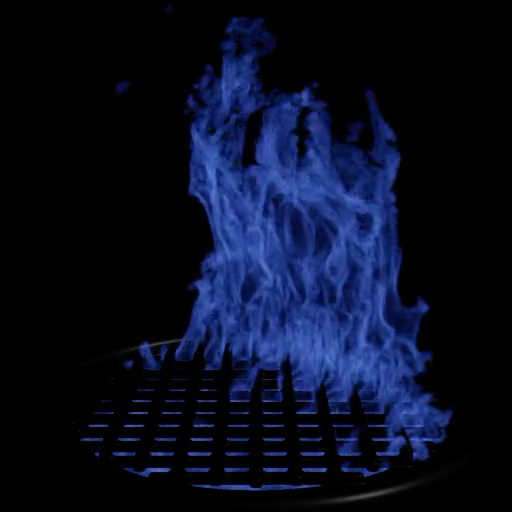}\includegraphics[height=.9in]{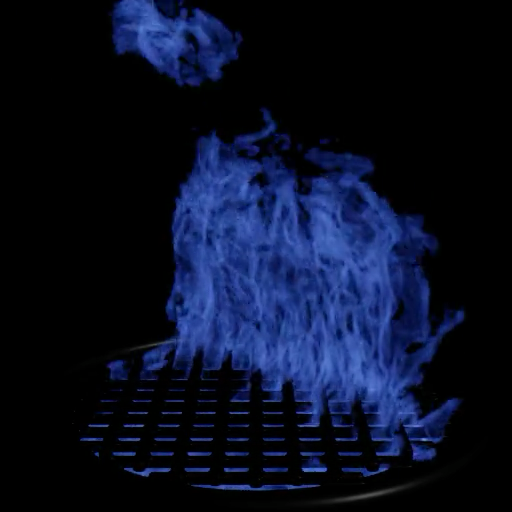}\includegraphics[height=.9in]{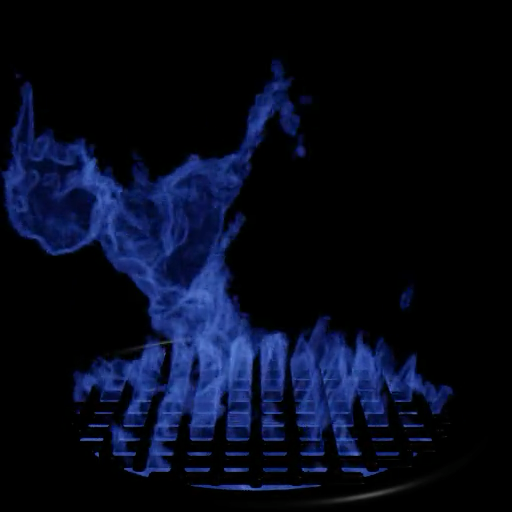}
    \caption{Left: Flame interface narrow-band. Right: Dynamic fire sim}
\vspace*{-.1in}
\label{fig:fire}
\end{figure}

We include micro-benchmarks from physics-based simulations of fire (Figure~\ref{fig:fire}), basing the combustion simulation on the techniques detailed in \cite{10.1145/566654.566643, NGUYEN200171, 10.1145/1275808.1276436}, which benefit from our novel Voxel Block Manager and efficient topological operators. The ghost fluid method employs a narrow-band level set representation—a truncated signed distance field (SDF)—to track the combustion flame front, or ``blue core''. This approach treats the combustion region as a discrete interface across which chemical properties exhibit jump conditions. In our implementation, the interface between fuel and air is rasterized into the NanoVDB sparse volumetric field, and dilated outwards from the surface crossing voxels to create a narrow band of active voxels. The Eikonal equation is iteratively solved over just this sparse narrow band using a stencil operation that visits $\pm$3 voxels in each of the 3 cardinal directions, for a total of 18 adjacent voxel visits in addition to the one central voxel visit. The discretization of the Euler equation is achieved with the optimally fifth-order accurate WENO scheme \cite{LIU1994200}, which is both computationally and arithmetically complex.

\paragraph{Level set tracking} The flame front is represented implicitly as the zero contour of a signed-distance field $\phi$, stored on a narrow band of active voxels around the interface. Each timestep advects $\phi$ by the flame-front velocity (derived from first-order detonation shock dynamics (DSD)
\cite{10.1145/1275808.1276436}), then reinitializes it back to a signed-distance function so the contour itself stays sharp while $\phi$ remains usable elsewhere as a smooth distance. Because the field is sparse, the active set has to move with the interface: each step (i) dilates the band so the next advect has source voxels to read from, (ii) semi-Lagrangian advects $\phi$ along $w$, (iii) reinitializes $\phi$ inside the band with a few iterations of a WENO-discretized Hamilton–Jacobi update \cite{LIU1994200}, and (iv) prunes voxels whose $|\phi|$ has grown past a band-width threshold. The microbenchmarks below isolate three operators from this pipeline—reinit, dilate, and prune.

\paragraph{Reinitialization micro-benchmark} The reinit kernel implements WENO5 on top of the new Voxel Block Manager (VBM) interface, and is compared against an accessor-only baseline that represents best-practice NanoVDB usage today. Each WENO5 update consumes a 19-point cross stencil per active voxel (center, plus $R{=}3$ taps along each of $\pm x$, $\pm y$, $\pm z$); the two implementations differ in how they fetch those off-center taps. The sparse grid stores active voxels in $8^3$ tiles, so an off-center tap usually lands in the same tile as the center but  occasionally crosses into a neighboring one. The accessor variant launches one CUDA block per tile whose threads walk the tree for every off-center tap, employing a small tree-walk helper that caches the last-touched tile, so consecutive within-tile taps are O(1) but every tile-boundary tap re-walks the tree. The VBM variant amortizes that bookkeeping across a 128-thread cluster of consecutive active voxels: the cluster cooperatively resolves its 128 tile addresses into shared memory once, and because $R{=}3$ is strictly less than the 8-voxel tile width, a single neighboring-tile pointer per axis covers every off-center tap any thread in the cluster can reach. The 19 stencil reads then compile to direct array loads against four resident tile pointers, with no per-tap tree walk.  Figure~\ref{fig:weno_sparsity} plots the throughput of the VBM-based WENO5 iteration against the accessor baseline as a function of mean leaf occupancy. VBM's throughput is essentially constant and up to $1.98\times$ that of the accessor implementation, whose throughput decreases as occupancy decreases and a growing fraction of its threads land on inactive slots.

\begin{figure*}[t]
    \centering
    \includegraphics[width=\textwidth]{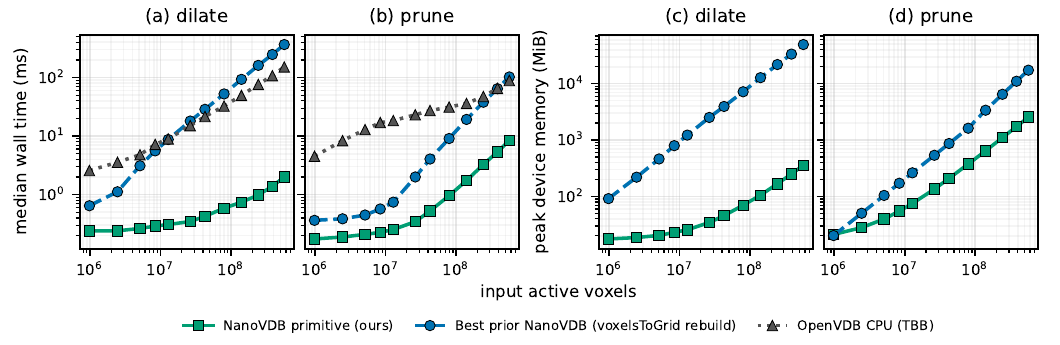}
\caption{
  Dilate (26-connected) and Prune ($|\phi| \le 1$) against the best prior NanoVDB route---rebuild from active coordinates through \texttt{voxelsToGrid}---and OpenVDB's TBB-parallel host implementation. Both operators run on identical input topologies: a flame front level set at 12 resolutions, $128\times192\times128$ to $1792\times2688\times1792$, 0.98\,M to 567\,M active voxels. Axes are logarithmic, so the margins are wider than they look. At the largest input dilate is $183\times$ faster than the rebuild and $76\times$ faster than the host, in $139\times$ less device memory (356\,MiB against 48.2\,GiB); prune is $12.1\times$ and $10.6\times$ faster, in $7\times$ less (2.49\,GiB against 16.9\,GiB). (a), (b): median wall time per call. (c), (d): peak device memory (the omitted host series has no device allocation).}
    \vspace*{-.1in}
    \label{fig:dilate_prune}
\end{figure*}

\paragraph{Topological operators micro-benchmark} The fluid simulation workload leverages the newly introduced \texttt{DilateGrid} and \texttt{PruneGrid} dynamic topology operators. In the level set pipeline dilate grows the active set by one voxel each step (we use the 26-neighbor face-edge-vertex pattern, so the advect's trilinear backtrace always lands on a source voxel), and prune is the inverse, dropping voxels whose $|\phi|$ has grown past the band-width threshold so the band can't grow unbounded across time steps. As Figure~\ref{fig:dilate_prune} shows, NanoVDB dilates a $567$\,M active-voxel topology in $2.0$\,ms using a pure topological mutation that touches no field storage. To match this throughput, a 32-bit field-carrying equivalent (OpenVDB) would have to read and rewrite $4(N_{in} + N_{out}) \approx 4.7$\,GB of field data. At the GPU's $1792$\,GB/s memory peak (Appendix~\ref{appendix:hardware-config}) that traffic alone takes $2.63$\,ms---longer than the $2.0$\,ms the topological formulation actually takes, so no field-carrying implementation can match it at any efficiency. Prune completes the same grid in $8.4$\,ms; the bandwidth argument does not extend to it, as its benchmark includes the predicate-eval read of the underlying field. Importantly, we compare dilation against a \texttt{voxelsToGrid}-based rebuild that drops every stencil tap that falls on an already-active voxel---about $13\times$ faster than rebuilding without tap elision, which itself runs out of memory above $69.5$\,M active voxels. Even so, our dilation is $183\times$ faster and peaks at $356$\,MiB against the baseline's $48.2$\,GiB of device memory, a $139\times$ difference and half the card's memory.

\begin{figure*}[t]
\includegraphics[width=\textwidth]{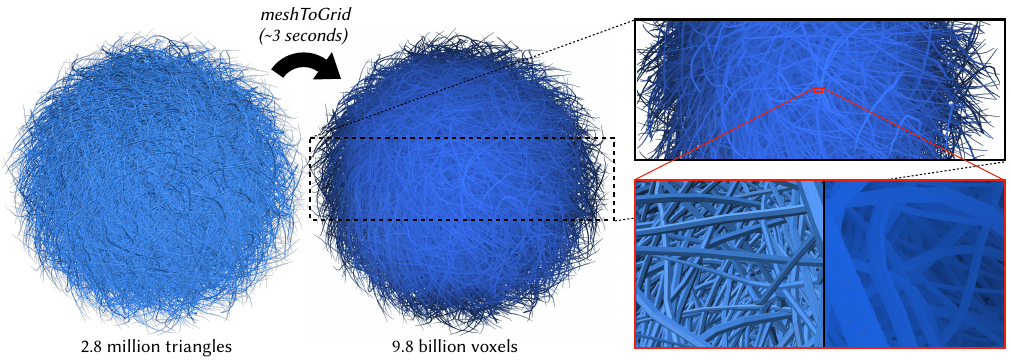}
\caption{Large-scale NanoVDB voxelization of a 2.8M-triangle hairball mesh \citep{McGuire2017Data}. At voxel size $0.001$ and narrow-band half-width $2$, our \texttt{MeshToGrid} implementation generates a sparse index grid with $9.88$B active voxels and UDF values stored in a sidecar in $3.24$s; the result is rendered directly with NanoVDB's hierarchical digital differential analyzer (HDDA). Hardware details in Appendix \ref{appendix:hardware-config}.}
\label{fig:meshToGrid}
\end{figure*}

\subsection{Large-Scale Shape Voxelization}\label{sec:large_udf}

We demonstrate one further application of our topological  pipeline: transforming a triangle mesh into a rasterized, sparse Unsigned Distance Field (UDF). Although distinct from the topological operators of \cref{sec:topology}, it constructively interweaves with the pipeline of \cref{fig:pipeline}. Since our objective is a narrow-band UDF, we use \emph{intersection} between a triangle and a volume primitive to indicate intersection with a thickness equal to the narrow band half-width.

We first stream through the triangle list, identifying how many root tiles ($4096^3$ voxels each) each triangle interacts with. Deduplicating these root tiles seeds steps 1--2 of the topological pipeline. A hierarchical broad-phase intersection progressively refines the \emph{(triangle,rootTile)} pairs into \emph{(triangle,leafNode)} pairs, where each leafNode spans $8^3$ voxels. This list marks the lower and leaf node bitmaps in Step 3; steps 4--5 then build the sparse tree that will hold the voxelized representation, with no voxels yet marked active. Step 6 visits the $8^3$ voxels of each leaf appearing in a \emph{(triangle,leafNode)} pair with a 512-thread CUDA block, marking active every voxel within the narrow-band half-width of the triangle. With the topology constructed, we allocate the UDF sidecar at exactly the storage the narrow band requires. Finally, we resolve the \emph{(triangle,leafNode)} list once more and perform a \emph{minimum reduction} in the sidecar, preserving at every voxel the smallest distance across all nearby triangles.

We evaluate the construction on the intricate ``hairball'' triangle mesh \citep{McGuire2017Data} shown in \cref{fig:meshToGrid}, consisting of 2.8 million triangles. The mesh is rasterized into a NanoVDB index grid with UDF values stored  in a separate sidecar array. At voxel size $0.003$ and narrow-band half-width $2$, OpenVDB's \texttt{meshToUDF} produced $1.025 \times 10^9$ active voxels in $23.87$ seconds, while our \texttt{MeshToGrid} implementation produced $1.026 \times 10^9$ active voxels in $2.30$ seconds---a roughly $10\times$ speedup at comparable resolution. At voxel size $0.001$ with the same narrow-band width, our implementation generated $9.88 \times 10^9$ active voxels in $3.24$ seconds; OpenVDB did not complete, terminating on memory exhaustion. These results highlight the ability of NanoVDB's dynamic sparse-grid construction to capture fine geometric detail at multi-billion-voxel scale while maintaining interactive construction times. For visualization, we render the sparse grid directly with the hierarchical digital differential analyzer (HDDA) in the  NanoVDB library~\cite{10.1145/2614106.2614136}. Rays traverse the sparse index-grid topology, and surface hits are identified by detecting UDF iso-crossings along the traversed active voxels, so the multi-billion-voxel result is visualized without conversion to a dense representation.

\section{Limitations and Future Work}

\subsubsection*{Topological Operations}
Although our topological operation framework can easily accommodate larger dilation stencils (as long as their radius does not exceed the size of leaf nodes, i.e~8 voxels, to avoid modifications to other ``generic'' components of the pipeline), stencils that have substantially more irregular (non-rectangular, non-symmetric) shapes might offer fewer opportunities for the aggressive bit-parallel optimizations we employed. We note however that \emph{box} stencils of radius up to 8 can be trivially accommodated with bit-parallel operations. Although we have included details on the implementation of morphological dilation, we have yet to implement its inverse operator, namely, morphological erosion. We plan to achieve this in terms of the existing prune operator. We should mention that our scratchpad-based bookkeeping commits to allocating approximately 16MB of storage for every $4096^3$ root tile that potentially overlaps with the output of our operation. Although it is unlikely that this limitation will prevent us from scaling up (the data payload would be substantially larger), it could potentially limit our performance advantage against reconstruction-from-scratch approaches for domains with very small size.

\subsubsection*{Voxel-sequential processing}
The key benefit of the Voxel-sequential access paradigm (via VBM) is maximization of compute bandwidth, by ensuring that data-parallel processors (e.g. GPUs) saturate execution units with regular work on \emph{active} voxels. This benefit is diminished in domains where leaf nodes are mostly dense, and those are also examples of applications where ``traditional'' NanoVDB grids (leaf-level dense) might be more appropriate than sparse Index-Grids. The benefit of maximizing compute density and saturation is also more prominent on compute bound, stencil-based operations; the WENO stencil in our fluid simulation examples, or sparse convolutions with deep vector-valued features (reducing to GEMM operations) would be best examples of such compute-bound workloads that stand to gain the most out of VBM. Memory-bound workloads (e.g. very small stencils on scalar fields, like a gradient or divergence Finite Difference operator) would struggle to see substantial benefits. Finally, workloads that access sparse data in spatially coherent (but not streaming/sequential) patterns, would remain closer to random-access tools (e.g. accessors) as the paradigm of choice.

\subsubsection*{Cross-resolution operators}
{}
The coarsening and refinement operators share the same processing pipeline and building blocks as dilation, merging, and pruning; nevertheless, they are not exercised end-to-end by any of the applications of \cref{sec:applications}. We also acknowledge that on an application-specific basis they may need to be paired with additional sidecar manipulation tools, beyond what we presented here. For example, one possible application would be the construction of a pyramid of sparse grids for multiresolution solvers, in the spirit of \citet{10.1145/2661229.2661269}. Such a workflow would also require a cross-resolution analogue of \texttt{inject}: whereas our present sidecar injection relates voxels across two topologies by coordinate identity, a 2:1 transfer is a reduction in one direction and an interpolation in the other, calling for an upsample/downsample (prolongation/restriction) operator optimized for sparse GPU computation. We defer these application-specific extensions to future work.
{}

\subsection*{Code Availability}
Our improvements to NanoVDB and the VoxelBlockManager implementation are made publicly available and open source as part of the Academy Software Foundation's OpenVDB Repository which can be found at \textit{\url{https://github.com/AcademySoftwareFoundation/openvdb}}. The code referenced in \cref{sec:reality_capture} and Appendix \ref{appendix:rc-full} are released in the same manner but as part of the fVDB Repository which can be found at \textit{\url{https://github.com/openvdb/fvdb-core}}.

\bibliographystyle{ACM-Reference-Format}
\bibliography{bibliography}

\clearpage
\appendix
\section{Hardware Configuration}\label{appendix:hardware-config}
All results, timings and memory usage numbers reported in our Applications, Section \ref{sec:applications}, were run on a machine running Ubuntu~26.04 using NVIDIA driver version 610.57 and CUDA~13.1 with an AMD Ryzen Threadripper PRO 3975WX 32-Cores CPU and an NVIDIA RTX PRO 6000 Blackwell Workstation Edition GPU with 96GB of GPU memory except for the results in the \textit{Large-Scale Shape Voxelization} Section \ref{sec:large_udf} which was done on a system with NVIDIA driver 595.84 using CUDA~13.2 and an AMD Ryzen Threadripper PRO 7965WX 24-Cores and an NVIDIA RTX PRO 6000 Blackwell Max-Q Workstation Edition GPU and the results in \textit{Perception and Reality Capture} Section \ref{sec:reality_capture}, including Appendix \ref{appendix:rc-full}, which was measured with an NVIDIA RTX 6000 Ada Generation GPU with 48GB of GPU memory.

\section{Reality Capture --- Full Evaluation}\label{appendix:rc-full}

This appendix expands the reality-capture summary of \cref{sec:reality_capture} and describes the remaining four algorithms: one-shot ESDF, incremental ESDF, log-odds occupancy, and dynamic-scene decay; the per-voxel-size cross-library evaluation; and the regimes in which each baseline outperforms our corresponding implementation.

\begin{figure}[h]
\centering
\includegraphics[width=\columnwidth]{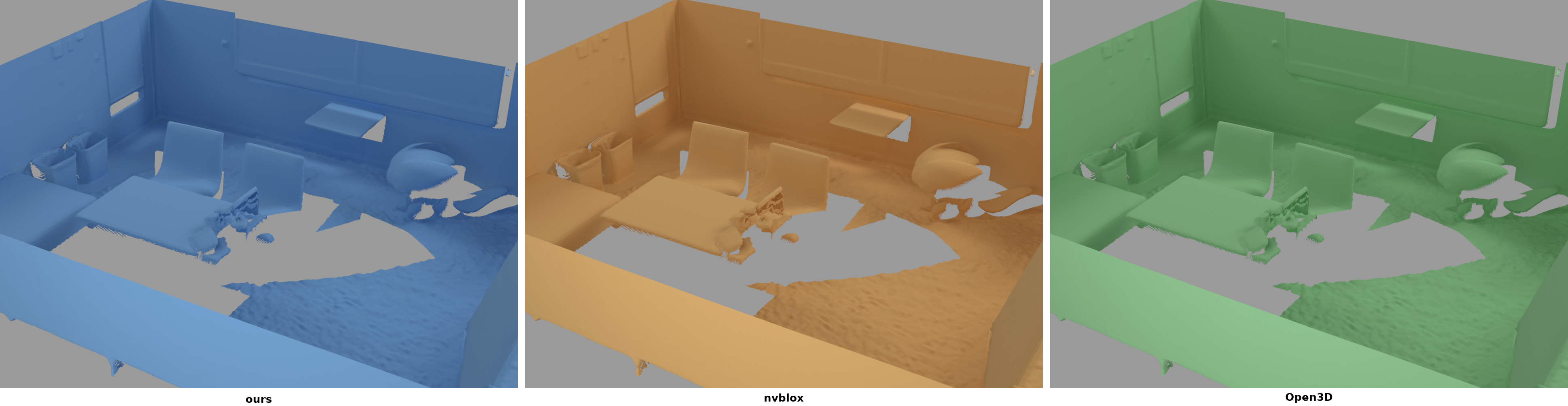}\\[2pt]
\includegraphics[width=\columnwidth]{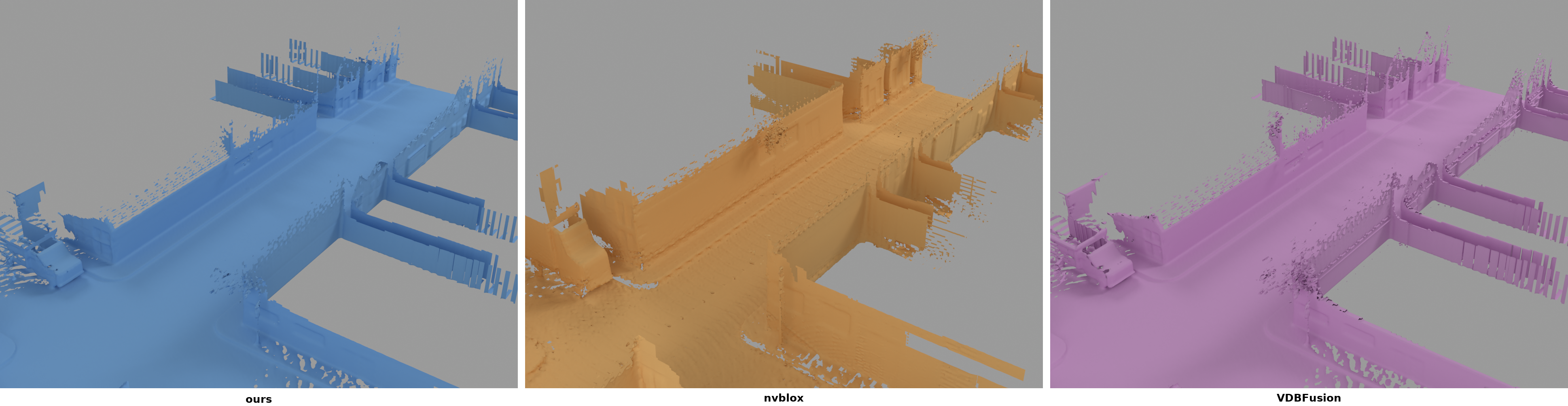}\\[2pt]
\includegraphics[width=\columnwidth]{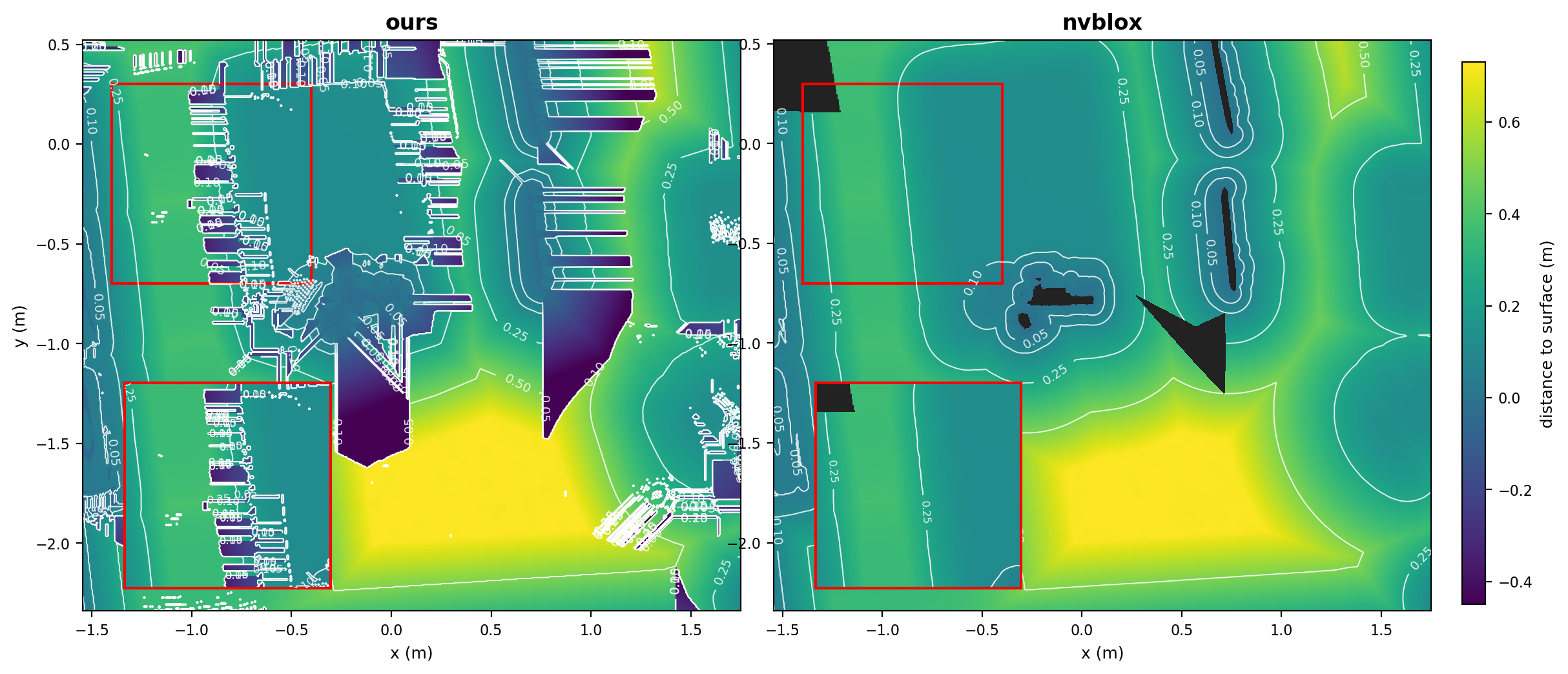}
\caption{Qualitative comparison on three of the reality-capture workloads. \emph{Top:} Replica \emph{office0} depth-TSDF mesh at 5\,mm (500 frames, ceiling clipped); nvblox's $8^3$-block padding shows as thicker walls. \emph{Middle:} Mai City seq 00 LiDAR-TSDF mesh at 10\,cm on a 60$\times$40\,m chunk; ours and VDBFusion show the cleanest sidewalk and lamp-post detail. \emph{Bottom:} Replica \emph{office0} ESDF horizontal slice at $z{=}-0.4$\,m (5\,mm voxels, $R{=}1$\,m) with iso-contours at 0.05 / 0.10 / 0.25 / 0.50\,m; the zoom inset (red box) highlights the resolution gap.}
\label{fig:rc-comparisons}
\end{figure}

\subsection{Algorithm overview}\label{sec:rc-algorithms}

\paragraph{Composition pattern.}
Every integrator we implement follows the same three-phase recipe per frame: \emph{(i)} build a topology shell of voxels that lie within the truncation band of a new observation and merge it into a persistent grid via \texttt{voxelsToGrid}~\cite{Williams:2024:fVDB} and \texttt{mergeGrids}~(\cref{sec:topology}); \emph{(ii)} re-align all per-voxel sidecar tensors to the new grid via \texttt{inject}~(\cref{sec:sidecar-ops}), with newly created voxels receiving a configurable initial value; \emph{(iii)} run an integrator-specific update kernel over the active voxel sequence. Update rules differ per integrator (weighted running average for TSDF, atomic log-odds add for occupancy, monotone-min sweep for ESDF), but the surrounding grow-then-inject structure is shared.

\paragraph{Parallelism strategies.}
The five algorithms considered in \cref{sec:reality_capture} and this appendix use three distinct GPU parallelism strategies depending on which is best matched to the workload's data flow. \emph{(i)}~Per-leaf-slot dispatch with an active-voxel mask, used by the depth-TSDF update kernel of \cref{alg:tsdf}: this is the simplest pattern and is competitive when leaf occupancy is high. \emph{(ii)}~Per-ray hierarchical-DDA, used by the LiDAR-TSDF update of \cref{alg:tsdf} and by \cref{alg:occupancy}: one thread per input point walks the active voxels along its ray and writes via lock-free atomic accumulation. \emph{(iii)}~The voxel-sequential VBM dispatch of \cref{sec:vbm}, used by the ESDF sweep of \cref{alg:esdf-cold}: each active voxel is mapped to a single thread independently of leaf occupancy, which is the right pattern for low-occupancy stencil work. We quantify the VBM speedup over per-leaf-slot dispatch on the ESDF sweep in \cref{sec:rc-limitations}.

\paragraph{Euclidean signed distance field.}
Given a TSDF on grid $G_{\mathrm{tsdf}}$ (\cref{alg:tsdf}), we compute the ESDF as a composition of a topology dilation (allocating the support band around the zero-crossing), a 26-neighbour monotone-min Jacobi sweep, and an optional prune of unreached frontier voxels. \cref{alg:esdf-cold} is a one-shot cold-start computation, and \cref{alg:esdf-warm} is an incremental warm-start variant. The fixed-point loop is the standard parallel-chamfer envelope for 26-connectivity used by FIESTA~\cite{han2019fiesta} and nvblox~\cite{millane2024nvblox}.

\begin{algorithm}[h]
\DontPrintSemicolon
\KwIn{TSDF grid $G_{\mathrm{tsdf}}$ with sidecars $d, w$;
       truncation $\tau$; max ESDF distance $R$.}
\KwOut{ESDF grid $G_{\mathrm{esdf}}$ and signed-distance sidecar $e$.}
$G_{\mathrm{esdf}} \leftarrow \textsc{dilateGrid}\!\left(G_{\mathrm{tsdf}},\,\lceil R/\tau\rceil\right)$\;
$e \leftarrow \textsc{seedKernel}(G_{\mathrm{esdf}}, d, w)$
   \tcp*{$e[v]=\operatorname{sign}(d[v])\!\cdot\!|d[v]|$ if $w[v]\!>\!0\,\wedge\,|d[v]|\!<\!\tau$, else $+\infty$}
$\mathit{changed} \leftarrow \mathit{True}$\;
\While{$\mathit{changed}$}{
    $\mathit{changed} \leftarrow \mathit{False}$\;
    \ForEach{$v \in G_{\mathrm{esdf}}$ \textnormal{iterated by VBM in parallel}}{
        $e_{\mathrm{new}} \leftarrow \min_{n \in \mathcal{N}_{26}(v)}\bigl(e[n] + \operatorname{sign}(e[n])\,\lVert x_v - x_n\rVert\bigr)$\;
        \If{$|e_{\mathrm{new}}| < |e[v]|$}{
            $e[v] \leftarrow e_{\mathrm{new}}$;\quad $\mathit{changed} \leftarrow \mathit{True}$\;
        }
    }
}
$e \leftarrow \textsc{clamp}(e,\,-R,\,+R)$\;
\Return $G_{\mathrm{esdf}}, e$\;
\caption{One-shot Euclidean signed distance field.}
\label{alg:esdf-cold}
\end{algorithm}

\begin{algorithm}[h]
\DontPrintSemicolon
\KwIn{Current TSDF $G_{\mathrm{tsdf}}, d, w$; previous ESDF $G_{\mathrm{prev}}, e_{\mathrm{prev}}$;
       optional dirty mask $m$ of length $|G_{\mathrm{tsdf}}|$; truncation $\tau$; max distance $R$.}
\KwOut{$G_{\mathrm{esdf}}, e$.}
\If{$m$ \textnormal{is given and} $m.\textsc{any}() = \mathit{False}$}{
    \Return $G_{\mathrm{prev}}, e_{\mathrm{prev}}$\tcp*{Python short-circuit}
}
$G_{\mathrm{esdf}} \leftarrow \textsc{dilateGrid}(G_{\mathrm{tsdf}},\,\lceil R/\tau\rceil)$\;
$e \leftarrow \textsc{inject}(G_{\mathrm{prev}} \to G_{\mathrm{esdf}},\,e_{\mathrm{prev}},\,\mathit{init}{=}{+}\infty)$\;
\ForEach{$v \in G_{\mathrm{esdf}}$ \textnormal{where} $(m \text{ is None or } m[v])$}{
    $e[v] \leftarrow \textsc{seedValue}(v, d, w)$\;
}
Run the fixed-point sweep of \cref{alg:esdf-cold} to convergence\;
\Return $G_{\mathrm{esdf}}, e$\;
\caption{Incremental ESDF with dirty-mask short-circuit.}
\label{alg:esdf-warm}
\end{algorithm}

\paragraph{Log-odds occupancy.}
Occupancy mapping (\cref{alg:occupancy}) shares \cref{alg:tsdf}'s grow-then-inject scaffold and replaces only the per-voxel update rule following OctoMap~\cite{hornung2013octomap}. Occupancy is the smallest of our integrators ($\sim$430 lines in one custom kernel).

\begin{algorithm}[h]
\DontPrintSemicolon
\KwIn{Grid $G$; log-odds sidecar $\ell$ of length $|G|$;
       $\Pi, Z, \tau$ as in \cref{alg:tsdf};
       $\ell_{\mathrm{hit}}, \ell_{\mathrm{miss}}$ Bayesian increments;
       clamp band $[\ell_{\min}, \ell_{\max}]$.}
\KwOut{$G', \ell'$.}
$\mathit{shell} \leftarrow \textsc{voxelsToGrid}\!\left(\textsc{LiDAREndpoints}(Z)\oplus\textsc{ball}(\tau)\right)$\;
$G' \leftarrow \textsc{mergeGrids}(G, \mathit{shell})$\;
$\ell' \leftarrow \textsc{inject}(G \to G', \ell, \mathit{init}{=}0)$\;
\ForEach{ray $r \in Z$ \textnormal{in parallel via HDDA ray-walk}}{
    \ForEach{voxel $v$ \textnormal{traversed by} $r$}{
        $\delta \leftarrow r.\textsc{isEndpoint}(v) \,?\, \ell_{\mathrm{hit}} : \ell_{\mathrm{miss}}$\;
        $\textsc{atomicAdd}(\ell'[v],\,\delta)$\;
    }
}
$\ell' \leftarrow \textsc{clamp}(\ell',\,\ell_{\min},\,\ell_{\max})$\;
\Return $G', \ell'$\;
\caption{Log-odds occupancy integration (single frame).}
\label{alg:occupancy}
\end{algorithm}

\paragraph{Dynamic-scene decay.}
Dynamic-scene support --- fading old observations so that moving objects stop persisting --- is a pure composition of two operators with no integrator-side code (\cref{alg:decay}, $\sim$30 lines of Python). On the first 500 frames of KITTI seq 00 at 10\,cm voxels with decay applied every 25 frames ($\gamma{=}0.5$, threshold $0.05$), the persistent active voxel count drops from 48.6\,M (no-decay baseline) to 10.6\,M --- a \textbf{78\% reduction} --- and is bounded at $\sim$10\,M voxels through the integration window, while the no-decay baseline grows monotonically (15\,M / 23\,M / 31\,M / 41\,M / 49\,M at 100 / 200 / 300 / 400 / 500 frames). The bounded state is the architectural effect of expressing decay as a tensor op on the weight sidecar plus a topology \texttt{pruneGrid}, with no integrator-side code aware of temporal coherence.

\begin{algorithm}[h]
\DontPrintSemicolon
\KwIn{Grid $G$; primary sidecar $s$; additional sidecars $s_{\mathrm{extra}}$;
       decay factor $\gamma\!\in\!(0,1)$; prune threshold $\varepsilon$.}
\KwOut{$G', s', s_{\mathrm{extra}}'$.}
$s \leftarrow s \cdot \gamma$\tcp*{tensor multiply}
$m \leftarrow s \geq \varepsilon$\tcp*{tensor mask}
$G' \leftarrow \textsc{pruneGrid}(G, m)$\;
$s' \leftarrow \textsc{inject}(G \to G', s)$\;
$s_{\mathrm{extra}}' \leftarrow \textsc{inject}(G \to G', s_{\mathrm{extra}})$\;
\Return $G', s', s_{\mathrm{extra}}'$\;
\caption{Dynamic-scene decay (per-tick).}
\label{alg:decay}
\end{algorithm}

\subsection{Cross-library evaluation}\label{sec:rc-eval}

\paragraph{Setup.}
We benchmark against nvblox~\cite{millane2024nvblox} for LiDAR/depth TSDF, ESDF, and occupancy; Open3D~\cite{zhou2018open3d} for depth TSDF; and VDBFusion~\cite{vizzo2022vdbfusion} for LiDAR TSDF. All measurements are on a single NVIDIA RTX~6000 Ada Generation GPU (48~GB) with PyTorch's \texttt{expandable\_segments} allocator. All systems use truncation distance $\tau = 3 \times \text{voxel size}$ for TSDF and maximum ESDF distance $R = 10 \times \text{voxel size}$, applied uniformly across baselines so the comparison is on integrator behaviour rather than parameter choices. nvblox in particular is configured with \texttt{projective\_integrator\_} \texttt{truncation\_distance\_vox} = 3.0 (matching $\tau$) and a  value \texttt{lidar\_projective\_integrator\_max\_} \texttt{integration\_distance\_m} of 60\,m on LiDAR plus an 8\,m equivalent on Replica/7-Scenes depth, in place of nvblox's upstream defaults of 4.0$\times$ truncation and 10\,m / 7\,m integration ranges; the upstream defaults clip nvblox's input to a $\pm$10\,m strip on LiDAR and inflate its truncation band by 33\%, neither of which is consistent with the workload fvdb processes.

\paragraph{Sensitivity to baseline defaults.}
As discussed, larger parameter values were chosen in order to produce larger scale scenes, consistent with the size of workloads our system was designed to process, rather than the upstream defaults assigned in nvblox.  The voxel-size thresholds which produce out-of-memory errors for nvblox in \cref{tab:occupancy} and \cref{tab:lidar-esdf} which support our scale claims are insensitive to using nvblox's default or the 6$\times$ larger value we chose.

We repeated the \cref{tab:occupancy} and \cref{tab:lidar-esdf} sweeps with nvblox at its upstream defaults: 10\,m LiDAR max integration distance and 4$\times$ voxel truncation (\texttt{ProjectiveIntegratorParams} defaults in nvblox version 0.0.10). In this experiment (\cref{tab:defaults-sensitivity}), nvblox's OOM voxel size moved by at most one voxel-size step in the sweep, and on Mai City it did not move at all.

\begin{table*}[h]
\centering
\small
\setlength{\tabcolsep}{3pt}
\begin{threeparttable}
\caption{\textbf{Sensitivity of scaling results} to chosen values of (LiDAR max integration distance, voxel truncation)}
\label{tab:defaults-sensitivity}

\begin{tabular}{l l l l}
\toprule
Workload &
nvblox @ (60\,m, $3\times$) &
nvblox @ defaults (10\,m, $4\times$) &
Ours @ defaults-matched \\
\midrule
Mai City TSDF (700 f) & OOM at 5\,cm &
OOM at 5\,cm\tnote{*} & 1\,cm (1.20\,B voxels, 45.9\,GB) \\
Mai City ESDF (700 f) & OOM at 5\,cm &
OOM at 5\,cm\tnote{*} & 2\,cm (30.0\,GB) \\
KITTI TSDF (seqs 00/02/05) & OOM at 10\,cm (all) &
\makecell[l]{%
  seq 02: OOM at 10\,cm\tnote{*};\\
  seqs 00/05: OOM at 5\,cm\tnote{\dag}%
} &
5\,cm / 5\,cm / 3\,cm \\
KITTI ESDF (100 f, 3 seqs) & OOM at 3\,cm &
OOM at 2\,cm\tnote{\dag} & 1\,cm (28--32\,GB) \\
\bottomrule
\end{tabular}

\begin{tablenotes}[flushleft]
\item[*] No change from the paper configuration.
\item[\dag] Improved by one resolution step.
\end{tablenotes}

\end{threeparttable}
\end{table*}

nvblox's OOM is triggered by its GPU block-hash rebuild (hard-coded \texttt{max\_load\_factor = 0.5}, \texttt{growth\_factor = 2}): when the accumulated map crosses half of the current power-of-two capacity, nvblox constructs a new table at twice the capacity while the old one is still resident and re-inserts every entry, with the backing voxel-block pool growing by the same $\times$2 policy. On Mai City at 5\,cm, this rebuild transiently demands \textasciitilde50 GB of device memory to admit 854 new blocks into the map. Using a 96 GB device, we measured that the run settles at 4.85 GB but peaks at 51.7 GB during the rebuild which will OOM with the 48 GB budget of the RTX~6000 Ada used in the experiments. Because the trigger is the accumulated block count crossing a capacity threshold, clipping the input from 60\,m to 10\,m reduces nvblox's steady-state memory \textasciitilde4$\times$ but does not keep the 700f Mai City trajectory below the fatal limit at 5\,cm. Our index-grid topology allocates exactly the observed voxels with no doubling rebuild, so peak $\approx$  steady state, which is why it fills 45.9 GB of the 48 GB budget effectively at 1\,cm on the same workload.
{}

\paragraph{Speed (LiDAR TSDF)}
On Mai City and KITTI we are 3.6--5.7$\times$ faster than nvblox at every cell where both run (\cref{tab:lidar-tsdf}), with the speedup narrowing monotonically as voxel size shrinks --- 5.7$\times$ at 40\,cm Mai City to 3.6$\times$ at 10\,cm Mai City; 4.8--5.3$\times$ at 40\,cm KITTI and 2.6--3.5$\times$ at 20\,cm KITTI. nvblox's per-block bookkeeping cost is roughly fixed per frame regardless of voxel size, while our cost scales with the count of voxels actually stored in the truncation shell, so the relative gap closes at finer voxels. On KITTI 10\,cm nvblox OOMs on every sequence (its hash allocator's resize step requests a contiguous slot buffer that the 48\,GB device can no longer supply), while ours runs at 12--21\,ms/frame. On the smallest KITTI sequence (seq 05, 2.2\,km) we additionally reach 5\,cm at 33.3\,GB of GPU memory with 726\,M voxels --- twice as fine as the longer sequences, which OOM there. VDBFusion (CPU OpenVDB)~\cite{vizzo2022vdbfusion} is 22--39$\times$ slower than ours across the KITTI 40--10\,cm sweep and 38--84$\times$ slower on Mai City across 40\,cm to 2\,cm; the cross-CPU gap is large enough that we report it for completeness only.

\paragraph{Speed (Occupancy).}
Occupancy is the cleanest sweep --- we win at every voxel size on KITTI (\cref{tab:occupancy}), with the speedup widening monotonically from \textbf{3.5$\times$ at 20\,cm to 11.5$\times$ at 1.5\,cm}, and continuing past nvblox's OOM threshold (1\,cm on every sequence) to \textbf{2.35\,B voxels in 44.1\,GB at 5\,mm}. The widening tracks the occupancy state's growing sparsity: nvblox's $8^3$ block padding caps its memory growth but penalises its time as the sparse occupancy sidecar stays sparse, while our per-voxel index stores exactly the active voxels and our update kernel touches only those.

\begin{table}[h]
\centering
\caption{\textbf{LiDAR TSDF on Mai City synthetic Velodyne sweeps and KITTI Odometry real-sensor LiDAR.} Per-frame integration time and peak GPU memory (CPU peak host RSS for VDBFusion); faster GPU time per row in bold. nvblox is configured with \texttt{lidar\_projective\_integrator\_max\_integration\_distance\_m} = 60\,m and \texttt{projective\_integrator\_truncation\_distance\_vox} = 3.0 to match our 60\,m-range LiDAR fusion and our 3$\times$ truncation; nvblox's upstream defaults of 10\,m and 4.0$\times$ would clip its input to a $\pm$10\,m strip and widen its band by 33\%. nvblox first OOMs at 5\,cm on Mai City and at 10\,cm on every KITTI sequence we tested. Ours scales through 1.25\,cm on Mai City at 37.1\,GB / 938.7\,M voxels and OOMs at 1.0\,cm; KITTI seq 00 and seq 02 OOM at 5\,cm while the smaller seq 05 reaches 5\,cm at 33.3\,GB on 726\,M voxels, and all three KITTI sequences OOM at 3\,cm. VDBFusion is 38--94$\times$ slower than ours on Mai City across 40--2\,cm and 22--41$\times$ slower on KITTI across 40--10\,cm; on KITTI 5\,cm VDBFusion runs cleanly on its CPU-side RAM where ours OOMs on the GPU, and on the seq 05 / 5\,cm cell where both run, VDBFusion is 27$\times$ slower. We did not run VDBFusion at KITTI 3\,cm: extrapolating from the seq 02 / 5\,cm cell's 96\,GB peak host RSS, the 3\,cm runs would exceed available system memory.}
\label{tab:lidar-tsdf}
\small
\setlength{\tabcolsep}{3pt}
\begin{tabular}{@{}llrrrrr@{}}
\toprule
                 &            & \multicolumn{2}{c}{Ours} & \multicolumn{2}{c}{nvblox} & VDBFusion \\
\cmidrule(lr){3-4}\cmidrule(lr){5-6}
Dataset / seq    & Voxel      & ms/f             & GB    & ms/f         & GB    & ms/f      \\
\midrule
\multirow{9}{*}{Mai City 700\,f}
                 & 40\,cm     & \textbf{3.44}    &  0.17 & 19.52        & 0.93  &  130.04   \\
                 & 20\,cm     & \textbf{3.80}    &  0.29 & 19.06        & 1.14  &  240.02   \\
                 & 10\,cm     & \textbf{5.07}    &  0.77 & 18.03        & 1.98  &  474.15   \\
                 &  5\,cm     & \textbf{11.22}   &  4.47 & OOM          & ---   &  965.28   \\
                 &  3\,cm     & \textbf{19.98}   &  9.30 & ---          & ---   & 1666.65   \\
                 &  2\,cm     & \textbf{32.26}   & 17.47 & ---          & ---   & 2723.86   \\
                 &  1.5\,cm   & \textbf{52.47}   & 27.48 & ---          & ---   &   ---     \\
                 & 1.25\,cm   & \textbf{68.10}   & 37.05 & ---          & ---   &   ---     \\
                 &  1.0\,cm   & OOM              & ---   & ---          & ---   &   ---     \\
\addlinespace[2pt]
\multirow{5}{*}{KITTI seq 00 (4541\,f)}
                 & 40\,cm     & \textbf{4.10}    &  1.23 & 21.07        & 1.14  &  118.86   \\
                 & 20\,cm     & \textbf{7.01}    &  4.02 & 19.21        & 3.05  &  224.62   \\
                 & 10\,cm     & \textbf{17.81}   & 15.78 & OOM          & ---   &  443.41   \\
                 &  5\,cm     & OOM              &  ---  & ---          & ---   &  965.39   \\
                 &  3\,cm     & OOM              &  ---  & ---          & ---   &  ---      \\
\addlinespace[2pt]
\multirow{5}{*}{KITTI seq 02 (4661\,f)}
                 & 40\,cm     & \textbf{4.38}    &  1.48 & 20.94        & 1.41  &  122.83   \\
                 & 20\,cm     & \textbf{7.88}    &  5.04 & 20.59        & 3.08  &  229.05   \\
                 & 10\,cm     & \textbf{21.49}   & 19.77 & OOM          & ---   &  462.97   \\
                 &  5\,cm     & OOM              &  ---  & ---          & ---   & 1003.33   \\
                 &  3\,cm     & OOM              &  ---  & ---          & ---   &  ---      \\
\addlinespace[2pt]
\multirow{5}{*}{KITTI seq 05 (2761\,f)}
                 & 40\,cm     & \textbf{3.92}    &  0.80 & 20.96        & 1.00  &  127.72   \\
                 & 20\,cm     & \textbf{5.80}    &  2.30 & 20.49        & 1.98  &  236.73   \\
                 & 10\,cm     & \textbf{12.00}   &  8.43 & OOM          & ---   &  469.41   \\
                 &  5\,cm     & \textbf{38.22}   & 33.31 & ---          & ---   & 1027.99   \\
                 &  3\,cm     & OOM              &  ---  & ---          & ---   &  ---      \\
\bottomrule
\end{tabular}
\end{table}

\begin{table}[h]
\centering
\caption{\textbf{KITTI Odometry occupancy mapping.} Per-voxel-size means over the three sequences \{00, 02, 05\}. nvblox is configured to match our 60\,m LiDAR range and 3$\times$ truncation (its defaults of 10\,m and 4.0$\times$ clip the input to a thin strip and widen the band). nvblox first OOMs at 1\,cm on every sequence (its hash allocator's resize step requests a single contiguous block of $\sim$2\,M slot entries that the 48\,GB device can no longer supply); ours scales to 2.35\,B active voxels in 44.1\,GB at seq 05 / 5\,mm. The split rows show finer voxels where ours continues running past nvblox's ceiling.}
\label{tab:occupancy}
\small
\setlength{\tabcolsep}{6pt}
\begin{tabular}{@{}lrrc@{}}
\toprule
Voxel                & Ours (ms/f) & nvblox (ms/f) & Speedup              \\
\midrule
20\,cm               &  5.8        &  20.1         & \textbf{3.5$\times$} \\
10\,cm               &  7.7        &  21.6         & \textbf{2.8$\times$} \\
 5\,cm               & 15.6        &  53.2         & \textbf{3.4$\times$} \\
 3\,cm               & 27.1        & 146.8         & \textbf{5.4$\times$} \\
 2\,cm               & 41.7        & 351.7         & \textbf{8.4$\times$} \\
1.5\,cm              & 59.1        & 678.6         & \textbf{11.5$\times$} \\
 1\,cm               & 93.3        & OOM           & ours only            \\
\addlinespace[2pt]
 8\,mm (seq 00 + 05) & 114.2       & OOM           & ours only            \\
 8\,mm (seq 02)      & 212.7       & OOM           & ours only            \\
\addlinespace[2pt]
 5\,mm (seq 00 + 05) & 270.2       & OOM           & ours only            \\
 5\,mm (seq 02)      & OOM         & OOM           & both OOM             \\
\bottomrule
\end{tabular}
\end{table}

\paragraph{Memory regime caveat.}
nvblox stores its state as fixed $8\!\times\!8\!\times\!8$ voxel blocks, allocating every cell in a block whether the cell is active or not; ours allocates exactly the active voxels with a per-voxel index. The two strategies trade places depending on how densely each $8^3$ block is occupied. On dense KITTI seq~02 / 20\,cm LiDAR TSDF, urban ground returns fill blocks nearly to capacity, so nvblox's per-block overhead amortises and our per-voxel index has the larger footprint: our peak GPU memory (5.04\,GB) exceeds nvblox's (3.08\,GB) by 1.6$\times$ (\cref{tab:lidar-tsdf}). The same regime appears at 40\,cm KITTI where the gap is small (within $\pm$20\%). Conversely, on KITTI Occupancy at 1.5\,cm where the state is spread thinly across $\sim$600\,M-voxel volumes, nvblox pays for many empty cells inside its blocks: our peak (3-seq mean 10.6\,GB) is 3.5$\times$ smaller than nvblox's (36.9\,GB) (\cref{tab:occupancy}), and at 1\,cm nvblox cannot fit at all while ours uses 19.8\,GB. The advantage holds asymptotically for sparsely-occupied workloads (the truncation-shell of depth or LiDAR TSDF below 10\,cm, the dilated band of ESDF, fine-voxel occupancy) and erodes for densely-occupied coarse-voxel ones. We report this rather than selecting datasets that suppress it.

\paragraph{Throughput scaling.}
On bounded-scene depth TSDF, Open3D's CUDA integrator~\cite{zhou2018open3d} is 11--46$\times$ faster than ours over the voxel range where both run on every scene (20--5\,mm), with the gap \emph{widening} monotonically because Open3D's per-frame block-hash bookkeeping is amortised over a tightly packed contiguous allocation while our cost grows with the count of voxels actually stored in the truncation shell. Open3D begins to fail on increasingly many Replica scenes as the voxel size shrinks below 5\,mm and crashes on every scene at 3\,mm (caption of \cref{tab:depth-tsdf-esdf}). Ours scales monotonically through 3\,mm at an 8-scene mean 44.50\,ms/f over $\sim$109\,M active TSDF voxels and an 8-scene-mean 1.61\,s ESDF cold sweep over 438\,M ESDF voxels. Our claims pertain to scale and breadth, not throughput parity on this single workload.

\begin{figure}[h]
\centering
\includegraphics[width=\columnwidth]{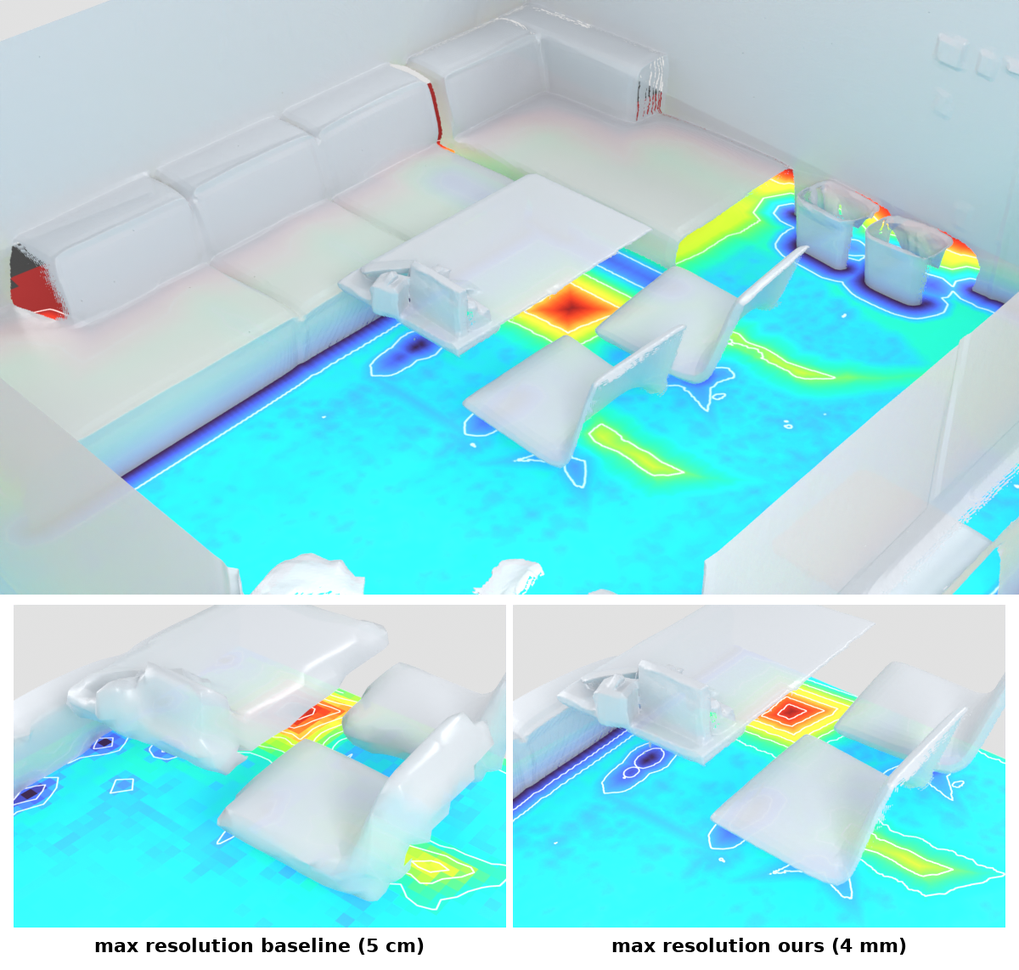}
\caption{The TSDF~+~ESDF pipeline pushed to the highest
    resolution each system can run on a single 48~GB GPU.
    $5$~cm is roughly the finest setting at which the block-hash TSDF
    systems we compare against (nvblox, Open3D, VDBFusion) remain
    stable on Replica scenes; NanoVDB's index-grid lets us reach $4$~mm
    ($12.5\times$ finer), at which point the mesh resolves individual
    chair backs and the ESDF iso-contours (drawn every $5$~cm of
    distance, in white) become smooth and dense rather than blocky
    and pixelated. }
\label{fig:rc-maxres}
\end{figure}

\begin{table}[h]
  \centering
  \footnotesize
\begin{threeparttable}
\caption{\textbf{Depth TSDF and ESDF on Replica (8 scenes $\times$ 2000 frames; mean) and 7-Scenes \emph{chess} (3000 frames).} Per-frame TSDF time and cold post-TSDF ESDF sweep time; fastest TSDF / ESDF per row in bold. Open3D and nvblox have no ESDF integrator on 7-Scenes in our setup; we ran nvblox only on Replica. nvblox is configured with \texttt{projective\_integrator\_max\_integration\_distance\_m} = 8\,m to match the depth ranges in Replica's NICE-SLAM rendering, slightly above its upstream default of 7\,m. nvblox fails with allocator OOM during its hash-resize step from Replica 5\,mm onward; Open3D fails with \texttt{illegal memory access} from its CUDA integrator on a staircase --- 8/8 Replica scenes succeed at 10\,mm and coarser, 6/8 at 5\,mm, 2/8 at 4\,mm (office0 and office1 only), 0/8 at 3\,mm. Ours scales monotonically through Replica 3\,mm and 7-Scenes 3\,mm with no per-frame creep across the 3000-frame chess trajectory (peak 10.6\,GB at chess 3\,mm). }
\label{tab:depth-tsdf-esdf}
\setlength{\tabcolsep}{3pt}
\begin{tabular*}{\columnwidth}{@{\extracolsep{\fill}}llrrrrr@{}}
\toprule
                 &           & \multicolumn{2}{c}{Ours} & \multicolumn{2}{c}{nvblox} & Open3D CUDA \\
\cmidrule(lr){3-4}\cmidrule(lr){5-6}
Dataset          & Voxel     & TSDF            & ESDF            & TSDF          & ESDF      & TSDF                \\
                 &           &      (ms/f)     &     (ms)        &       (ms/f)  &      (ms) &     (ms/f)               \\
\midrule
\multirow{5}{*}{\shortstack[l]{Replica\\$8\!\times\!2000$\,f}}
                 & 20\,mm    &  2.64           & \textbf{38.46}   & 0.81         &  75.60    & \textbf{0.25}             \\
                 & 10\,mm    &  5.23           & \textbf{125.83}  & 6.43         & 529.54    & \textbf{0.26}             \\
                 &  5\,mm    & \textbf{19.05}  & \textbf{617.71}  & OOM          & OOM       & \textbf{0.42}$^{\ddagger}$ \\
                 &  4\,mm    & \textbf{28.06}  & \textbf{926.85}  & OOM          & OOM       & \textbf{0.41}$^{\ddagger}$ \\
                 &  3\,mm    & \textbf{44.50}  & \textbf{1614.89} & OOM          & OOM       & crash                     \\
\addlinespace[2pt]
\multirow{4}{*}{\shortstack[l]{7-Scenes \\\emph{chess}\\3000\,f}}
                 & 20\,mm    &  1.96           & n/a              & n/a          & n/a       & \textbf{0.21}             \\
                 & 10\,mm    &  2.31           & n/a              & n/a          & n/a       & \textbf{0.23}             \\
                 &  5\,mm    &  5.61           & n/a              & n/a          & n/a       & \textbf{0.26}             \\
                 &  3\,mm    &  16.20          & n/a              & n/a          & n/a       & \textbf{0.33}             \\
\bottomrule
\end{tabular*}

\begin{tablenotes}[flushleft]
\item[$^{\ddagger}$] Open3D mean over the Replica scenes that succeed.
\end{tablenotes}
\end{threeparttable}
\end{table}

\paragraph{Long-trajectory persistence.}
To stress the grow-then-inject discipline of \cref{alg:tsdf} on an indoor sequence longer than Replica or KITTI, we integrate 7-Scenes \emph{chess}~\cite{shotton2013scenes} for its full 3000 RGB-D frames at 20/10/5/3\,mm depth TSDF (lower block of \cref{tab:depth-tsdf-esdf}). Ours reaches 3\,mm at 187\,M active voxels in 10.6\,GB with bounded mean per-frame cost across the trajectory; Open3D runs every cell at 9--49$\times$ our throughput, consistent with the Replica \emph{throughput scaling} above.

\paragraph{ESDF scaling.}
On Mai City synthetic LiDAR over the full 700-frame trajectory, nvblox's cold ESDF is faster than ours at 20\,cm by 1.4$\times$ but slower at 10\,cm by 1.8$\times$, and cannot build the TSDF substrate below 5\,cm (\cref{tab:lidar-tsdf}); we scale to \textbf{3\,cm with 1.28\,B ESDF voxels in 30.6\,GB} on the same trajectory, and OOM at 2\,cm (\cref{tab:lidar-esdf}). On KITTI we use a 100-frame TSDF prefix per sequence so that the dilated ESDF band fits in GPU memory across the voxel-size sweep on both systems; nvblox is 1.4--2.0$\times$ faster at 20\,cm and competitive at 10\,cm (KITTI seq 00: 0.88$\times$; seq 02: 1.17$\times$ slower; seq 05: 1.60$\times$ slower), then we are 1.8--1.9$\times$ faster at 5\,cm on seq 00 and seq 05, and nvblox OOMs from seq 02 / 5\,cm onward. What ours wins on KITTI is \emph{scale}: at 3\,cm on seq 02 we run at 3.93\,s on 1.20\,B active voxels (27.7\,GB) where nvblox is out of memory, and at 2\,cm on seq 00 we reach \textbf{1.67\,B active voxels in 38.6\,GB}. The dirty-mask short-circuit of \cref{alg:esdf-warm} reduces the no-change case to sub-400\,$\mu$s at every voxel size we measured.

\begin{table*}[ht]
\centering
\caption{\textbf{LiDAR ESDF on Mai City over the full 700-frame trajectory and on KITTI Odometry over a 100-frame TSDF prefix per sequence.} Active voxel count, cold ESDF time, dirty-mask idempotent time, and peak GPU memory. nvblox is configured to match our 60\,m LiDAR range and 3$\times$ truncation. nvblox first OOMs at Mai City 5\,cm (TSDF substrate fails) and at KITTI seq 02 / 5\,cm; on KITTI seq 00 and seq 05 nvblox runs at 5\,cm but OOMs at 3\,cm. Ours scales to 1.67\,B active voxels at 38.6\,GB on KITTI seq 00 / 2\,cm and 1.28\,B voxels at 30.6\,GB on Mai City 700-frame / 3\,cm.}
\label{tab:lidar-esdf}
\small
\begin{tabular}{@{}llrrrrrr@{}}
\toprule
                 &        &                  & \multicolumn{3}{c}{Ours} & \multicolumn{2}{c}{nvblox} \\
\cmidrule(lr){4-6}\cmidrule(lr){7-8}
Dataset / seq    & Voxel  & ESDF voxels      & Cold (ms) & Warm ($\mu$s)       & Peak (GB)       & Cold (ms) & Peak (GB) \\
\midrule
\multirow{5}{*}{Mai City 700\,f}
                 & 20\,cm & 38.7\,M          &   129.0   &  25\,$\mu$s         &  2.04           &   174.5   &  1.67 \\
                 & 10\,cm & 129.4\,M         &   416.4   &  32\,$\mu$s         &  4.17           &   740.4   &  2.50 \\
                 &  5\,cm & 459.2\,M         &  1628.0   &  59\,$\mu$s         & 11.82           &   OOM     &  ---  \\
                 &  3\,cm & \textbf{1.28\,B} &  5567.9   & 266\,$\mu$s         & \textbf{30.59}  &   OOM     &  ---  \\
                 &  2\,cm & OOM              &   ---     & ---                 & ---             &   OOM     &  ---  \\
\addlinespace[2pt]
\multirow{5}{*}{KITTI seq 00 (100\,f)}
                 & 20\,cm & 21.6\,M          &    81.3   &  25\,$\mu$s         &  0.66           &    41.4   &  1.46 \\
                 & 10\,cm & 76.3\,M          &   212.5   &  33\,$\mu$s         &  1.94           &   186.2   &  1.90 \\
                 &  5\,cm & 277\,M           &   800.4   &  49\,$\mu$s         &  6.61           &  1477.1   &  3.37 \\
                 &  3\,cm & 779\,M           &  2571     & 191\,$\mu$s         & 18.12           &   OOM     &  ---  \\
                 &  2\,cm & \textbf{1.67\,B} &  5906     & 365\,$\mu$s         & \textbf{38.61}  &   OOM     &  ---  \\
\addlinespace[2pt]
\multirow{5}{*}{KITTI seq 02 (100\,f)}
                 & 20\,cm & 27.2\,M          &    87.0   &  25\,$\mu$s         &  0.80           &    85.8   &  1.53 \\
                 & 10\,cm & 107\,M           &   327.5   &  33\,$\mu$s         &  2.66           &   383.9   &  2.41 \\
                 &  5\,cm & 416\,M           &  1268     &  66\,$\mu$s         &  9.83           &   OOM     &  ---  \\
                 &  3\,cm & 1.20\,B          &  4184     & 299\,$\mu$s         & 27.71           &   OOM     &  ---  \\
                 &  2\,cm & OOM              &   ---     & ---                 & ---             &   OOM     &  ---  \\
\addlinespace[2pt]
\multirow{4}{*}{KITTI seq 05 (100\,f)}
                 & 20\,cm & 19.5\,M          &    62.6   &  24\,$\mu$s         &  0.62           &    43.0   &  1.52 \\
                 & 10\,cm & 70.5\,M          &   218.5   &  33\,$\mu$s         &  1.81           &   349.4   &  1.88 \\
                 &  5\,cm & 258.9\,M         &   788.2   &  50\,$\mu$s         &  6.18           &  1516.6   &  3.34 \\
                 &  3\,cm & 732.3\,M         &  2504.9   & 183\,$\mu$s         & 17.07           &   OOM     &  ---  \\
\bottomrule
\end{tabular}
\end{table*}

\subsection{Performance Trade-offs and Limitations}\label{sec:rc-limitations}

Four regimes where the architectural argument degrades:

\begin{enumerate}
\item \textbf{Open3D wins bounded depth-TSDF throughput by 11--46$\times$ at 20--5\,mm Replica voxels} (see \emph{throughput scaling} above), with the gap widening monotonically. nvblox is also faster than ours at 20\,mm (3.3$\times$) but loses to ours from 10\,mm onward (\cref{tab:depth-tsdf-esdf}). Open3D's integrator has a decade of head start on hand-tuned block-hash allocators; we do not match it on this axis. We do scale past both: at 4\,mm nvblox fails on every scene and Open3D on 6\,/\,8; at 3\,mm both fail on every scene.

\item \textbf{Cold ESDF cost is 1.0--2.0$\times$ higher than nvblox's at 20\,cm KITTI voxels} (tied on seq 02, 1.5$\times$ slower on seq 05, 2.0$\times$ slower on seq 00), and 1.5$\times$ higher at 40\,cm Mai City. We become uniformly faster from 20\,cm on Mai City and from 5\,cm on KITTI; nvblox OOMs from 5\,cm KITTI seq 02 and 3\,cm everywhere else.

\item \textbf{Our memory advantage over nvblox's per-block storage is regime-dependent.} Dense urban LiDAR at coarse voxels (KITTI seq 02 / 20\,cm TSDF) flips the per-voxel comparison to nvblox-favourable by 1.6$\times$ since its $8^3$ blocks are nearly full; sparse-state fine-voxel occupancy (KITTI 1.5\,cm) flips it back to a 3.5$\times$ advantage in our favour, and at 1\,cm nvblox cannot fit at all. The advantage holds asymptotically for sparsely-occupied applications (\emph{Memory regime caveat}, above).

\item \textbf{The VBM speedup over per-leaf-slot dispatch on the ESDF sweep is 3--6\%, not the order-of-magnitude one might expect} from a hierarchical-dispatch-replacement story. ESDF leaves end up densely occupied by the dilation step (\cref{alg:esdf-cold}, line~1), so per-leaf-slot iteration wastes few threads on inactive entries. VBM's advantage is realised on low-occupancy stencil work where active voxels are sparse within each leaf.
\end{enumerate}

A fifth caveat was present in early designs of \cref{alg:esdf-warm}: incremental ESDF warm-reuse was 400$\times$ slower than nvblox's \texttt{update\_\allowbreak esdf}. The dirty-mask Python short-circuit of \cref{alg:esdf-warm} reduces the no-change case to sub-400\,$\mu$s and matches nvblox's sub-100\,$\mu$s block-cache hit on the typical sparse-update case.

\end{document}